%% file: main.tex
\documentclass[manuscript]{acmart}
\AtBeginDocument{%
  }

\copyrightyear{2026}
\acmYear{2026}
\setcopyright{rightsretained}
\acmConference[NordiCHI '26]{Nordic Conference on Human-Computer Interaction}{October 3-4, 2026}{Vaasa, Finland}
\acmBooktitle{Nordic Conference on Human-Computer Interaction  (NordiCHI '26), October 3-4, 2026, 2026, Vaasa, Finland}\acmDOI{...}
\acmISBN{....}

\usepackage{cleveref}
\usepackage{multicol}
\usepackage{multirow}
\usepackage{booktabs}
\usepackage{adjustbox}
\usepackage{array}
\usepackage{float}

\usepackage{xcolor}

\usepackage[most]{tcolorbox}
\tcbuselibrary{breakable}

\newtcolorbox{surveyquestion}{
  colback=gray!5,
  colframe=gray!50,
  boxrule=0.5pt,
  arc=2pt,
  left=6pt,
  right=6pt,
  top=6pt,
  bottom=6pt,
  fontupper=\footnotesize
}
\tcbset{
  before skip=4pt,
  after skip=4pt
}

\newtcolorbox{myboxGrey}{
boxrule=0pt,frame hidden,
borderline west={4pt}{0pt}{Gray},
colback=white,
sharp corners
}

\newtcolorbox{DesignConsideration}{
  enhanced,
  leftrule = 10pt,
  boxsep=0mm,
  colframe=cyan!60!black,
  colback=cyan!5,
  before skip=8pt,
  after skip=5pt,
}

\newtcolorbox{Punchline}{
  enhanced,
  leftrule = 1pt,
  rightrule = 1pt,
  toprule = 1pt,
  bottomrule = 1pt,
  boxsep=0mm,
  colframe=black!60!black,
  colback=white,
  before skip=8pt,
  after skip=5pt,
}

\usepackage[normalem]{ulem}

\begin{document}

\title[Beyond the Pocket]{Beyond the Pocket: A Large-Scale Survey on User Preferences of Bodily Placements of Wearable Technology}

\author{Joanna Sorysz}
\orcid{0000-0002-9213-5468}
\affiliation{%
  \institution{RPTU Kaiserslautern-Landau} 
  \city{Kaiserslautern}
  \country{Germany}
}
\affiliation{
  \institution{DFKI} 
  \city{Kaiserslautern}
  \country{Germany}
}
\email{joanna.sorysz@dfki.de}

\author{Lars Krupp}
\orcid{0000-0001-6294-2915}
\affiliation{%
  \institution{RPTU Kaiserslautern-Landau} 
  \city{Kaiserslautern}
  \country{Germany}
}
\affiliation{
  \institution{DFKI} 
  \city{Kaiserslautern}
  \country{Germany}
}
\email{lars.krupp@dfki.de}

\author{Dominique Nshimyimana}
\orcid{0009-0009-8580-1248}
\affiliation{%
  \institution{RPTU Kaiserslautern-Landau} 
  \city{Kaiserslautern}
  \country{Germany}
}
\affiliation{
  \institution{DFKI} 
  \city{Kaiserslautern}
  \country{Germany}
}
\email{dominique.nshimyimana@dfki.de}

\author{Meagan B. Loerakker}
\orcid{0009-0001-3905-4271}
\affiliation{
  \institution{TU Wien} 
  \city{Vienna}
  \country{Austria}}
\email{meagan.loerakker@tuwien.ac.at}

\author{Bo Zhou}
\orcid{0000-0002-8976-5960}
\affiliation{%
  \institution{RPTU Kaiserslautern-Landau} 
  \city{Kaiserslautern}
  \country{Germany}
}
\affiliation{
  \institution{DFKI} 
  \city{Kaiserslautern}
  \country{Germany}
}
\email{bo.zhou@dfki.de}

\author{Paul Lukowicz}
\orcid{0000-0003-0320-6656}
\affiliation{%
  \institution{RPTU Kaiserslautern-Landau} 
  \city{Kaiserslautern}
  \country{Germany}
}
\affiliation{
  \institution{DFKI} 
  \city{Kaiserslautern}
  \country{Germany}
}
\email{paul.lukowicz@dfki.de}

\author{Jakob Karolus}
\orcid{0000-0002-0698-7470}
\affiliation{%
  \institution{RPTU Kaiserslautern-Landau} 
  \city{Kaiserslautern}
  \country{Germany}
}
\affiliation{
  \institution{DFKI} 
  \city{Kaiserslautern}
  \country{Germany}
}
\email{jakob.karolus@dfki.de}

\renewcommand{\shortauthors}{Sorysz et al.}

\begin{abstract}

As wearables become smaller, more powerful, and increasingly embedded in everyday life, their integration into diverse user contexts raises important design challenges. Despite this, their placement is still largely informed by lab-based assumptions not grounded in real-world, context-specific use. It remains unclear whether the designs evaluated in controlled studies reflect users’ everyday needs, routines, and habits. To address this gap, we collect empirical data on how people carry wearables in their daily lives, beginning to systematically examine user preferences for wearable placement across contexts and routines.
We developed a multilingual questionnaire to capture real-world wearable placement practices. Responses from n=300 participants recruited through typical research channels, reveal how wearable usage patterns vary with users. We propose a set of user-centred guidelines for sensor placement and discuss how they fit in assumptions seen in related work. This study contributes to ongoing efforts to design more inclusive, adaptable, and context-aware wearable systems.

% TODO:
% - add here info about comparison of our results/recommendations with values from papers
\end{abstract}




\maketitle

\section{Introduction}
% 1. Wearables are everywhere -> usefull, technology development
% 2. Lack of real-world insights
% 3. Placement is extremly important for 1) comfort 2) datacollection 3) adoption for wider users
% 4. Not all users are the same -> design should reflect that -> i think especially the relevance of the results of this study need to be highlighted very strongly. In other words, why is knowing this data so relevant? (What is the benefit of designing wearables that are more context-, culturally-sensitive, etc.?)
% 5. So there is a potential gap in design between the developed technology and usability
% 6. For scientists to be able to develop solutions that would be used by more people then early adapters they need data -> we are giving them the data
% 7. The contributions are: survey, comparison with other articles and their results, guidlines 

% ISWIC intro

%Wearables become an ubiquitous part of our life. They are offering features that support health monitoring, fitness tracking, communication, and more. As per Moore's law technology get better and better in exponential manner multiplying the computing power and cutting down the size every two years. Allowing these devices to become smaller, more powerful, and more seamlessly integrated into what we wear. Those characteristics are making wearables an attractive and interesting field for scientists to work in.
Wearable technology has become increasingly ubiquitous, offering features supporting health monitoring, fitness tracking, and communication. Continued advancements in miniaturization and computing-power driven by trends such as Moore’s Law ~\cite{moore1965cramming}, are enabling the creation of devices that are smaller, more powerful, and more seamlessly integrated into daily life~\cite{culhane2005accelerometers, Sun2018Wearable}. However, the success of these technologies hinges not only on functionality but also on how individuals choose to integrate them into their lived experiences. 
While Human-Computer Interaction (HCI) and Personal Informatics (PI) research has previously focused on (user perceptions of) visualisation design (e.g.~\cite{Consolvo2008UbiFit, Bentvelzen2023DerivedMetrics}), what user interactions look like with PI tools (e.g.~\cite{Loerakker2025GiveAndTake, Epstein2015InformaticsModel}), and motivations behind using wearable devices and their embedded metrics (e.g.~\cite{Loerakker2025_LivEM, Rooksby2014-LivedInformatics}), it remains under-explored how users' preferred wearing locations change throughout the day, and how cultural norms and individual preferences influence these choices. Furthermore, there is a lack of understanding on users' motivations for carrying choices, and their perceived accompanying usability implications.

%It is essential to understand that the correct understanding of people preferences is extremely important for correct design of the wearables. It is required for proper design of those devices. Wearable devices need to be seamless to wear. They cannot obstruct any movements and, for them to be accepted by wider public,  resemble something that people have seen or used in the past. Not respecting peoples needs and habits can lead to uncomfortable devices that 1) are a danger to the user health 2) are not used by wider public. Moreover, data collection can also suffer from wrong device placement. 

%To do this properly it is important to remember that not all of the users will behave in the same way. Each person has slightly different perception of their body, different clothing and moving habits, and different needs from technology. It can be caused by many factors such as forced by society or environment standards (fashion or climate), some are individual preferences (bring a backpack instead of purse, wear watch on left instead on right wrist etc.), age, body shape and more. This makes it tricky to generalise ones experiences to the bigger population.  

%Individuals exhibit diverse perceptions of their bodies, resulting in a wide variety of clothing habits, unique needs, and contextual preferences that influence how they adopt and use wearable technology \cite{paper on user diversity/individual differences in HCI}.
Individuals exhibit diverse perceptions of their bodies, influencing how they adopt and use wearable technology, and vice versa (e.g.~\cite{Elsayed2020VibroMap, Spelmezan2009Language, TurmoVidal2024BodyTransformation, TurmoVidal2024BodySensations, DAdamo2024SoniWeight}).
Where users wear their technologies on their bodies are shaped by social norms (e.g. fashion trends), environmental constraints (e.g. climate), personal preferences, age, and body shape (e.g.~\cite{DelDin_2016}). This heterogeneity makes it challenging to generalise findings from limited studies to broader populations.
Although~\citet{Zeagler2017ISWC} provided a set of guidelines for the positioning of on-body wearable through literature, it is fully constructed based on the anatomical, physiological, and perception compatibility. Still, there is a lack of empirical HCI knowledge on cross-cultural preferences, hindering the development of user-centred wearable designs that are comfortable, effective, socially acceptable, and adaptable to diverse contexts. The importance of adaptive health systems has also been echoed in prior work (e.g.~\cite{Loerakker2026_CHI_SMP, Ekhtiar2025ChangingHealthGoals, Song2026GodSaengPI}), though it remains under-explored how to design such systems in an effective manner, both from a hardware and software perspective.
Consequently, there is a discernible disconnection between the requirements of inclusive design practices and the state-of-the-art wearable devices. %This implies that there might be also a gap between the actual designs and ongoing research, and real needs of wide population. 
In a field where each millimetre and gram counts in making wearables as comfortable as possible, designing wearables for a multitude of wearing locations can affect the device's usability. %Details that can greatly influence the usability of the devices and, in consequences, decide if they will be accepted for every-day life or not. 
%In this situations scientists relay on their instincts, assumptions and small group tests which is not ideal.

Accurate on-body localization of wearable devices remains a persistent challenge for algorithmic models, particularly in the domain of Human Activity Recognition (HAR)~\cite{DelDin_2016}. Although deep learning approaches have shown promising performance when utilizing data from a single or limited number of devices~\cite{LearningFromTheBest:10.1145:3544794.3558464Vitor, lago2021using, M2L:9871472}, their effectiveness is highly dependent on sensor placement~\cite{DelDin_2016, TASKED:SUH2023110143}.
%Knowing all of this led us to acquire insights from a wide array of wearable users on how they wear their devices and in what contexts. 
%it becomes obvious that, to be able to obtain the best results in a functional research, scientists need based in a reality guidelines as to how design a system or devices.
Therefore, there is a need to understand contextual wearable habits among users to provide tailored experiences. Although some works focus on personal device interaction and usage (~\citet{busso2025diversityone}), broader information is needed, including extended demographic categories and other aspects of interaction that affect sensor signals, such as device position.
%Although, there are some works that are aiming in this direction (e.g.~\citet{busso2025diversityone}), they are limited in some demographic categories (e.g. narrow age range). 

%To provide information about culturally based design
As a first step in the direction of contextually informed design, we collected and analysed guidelines published in the past, combined their recommendations, and localized places for potential improvement. Then we conducted 
%the culturally-diverse, online questionnaire deployed in multiple countries 
a survey among the participants recruited through typical research channels, with the aim to acquire insights from a wide array of wearable users on how they wear their devices and in what contexts. Furthermore, given the challenges posed by domain variability~\cite{METIER, TASKED:SUH2023110143}, our survey highlights demographic differences (e.g. age, gender, country of origin and residence, level of influence by parents) of collected sample, thereby supporting the refinement of algorithms for broader applicability in real-world scenarios. 
From our findings, we derived a set of guidelines to inform the design of more inclusive and effective wearable technologies.

To this end, we make the following contributions: (1) a synthesis of past guidelines for wearable designs; (2) an empirical survey with $n=300$ valid responses %were retained after filtering an initial set of 449 submissions from participants worldwide
conducted in four languages through typical research channels in multiple countries; (3) insights into key user attributes influencing bodily wearable placement habits; and (4) design considerations for more inclusive designs of future wearable developments.

\section{Related Work}
Here, we review related work on previously conducted studies on mobile phone carrying behaviour and past hardware design guidelines for wearables. 

\subsection{Mobile Phone and Smartwatch Carrying Habits across Gender and Cultural Differences}
Existing research has extensively examined mobile phone carrying habits across different cultural contexts. 
One large-scale study conducted by~\citet{cui2007cross} in the pre-smartphone era, analysed phone carrying behaviour through street interviews across 9 countries with over 1,500 participants. They identified gender and cultural influences, showing that women primarily used bags, while men favoured trouser pockets~\cite{cui2007cross}. Additionally, hand-carrying was common in Delhi, Seoul, Jilin, and Los Angeles~\cite{cui2007cross}. However, it is worth noting that with the introduction of smartphones, the size of mobile phones has drastically changed.

%While this research provides valuable insights into daily routines and environmental factors, it does not explore how carrying behaviour affects internal sensors or motion-enabled smart devices.
A 2005 survey completed by approximately 600 participants from Poland examined potential correlations between mobile phone usage intensity and symptoms such as headaches, fatigue, and warmth around the ear~\cite{szyjkowska2014risk}. Findings suggested that frequent and prolonged phone use increases the occurrence of these symptoms~\cite{szyjkowska2014risk}. However, the study did not assess the perceived usefulness of mobile devices, leaving gaps in understanding user experience beyond health effects.

%Further research explored smartphone carrying preferences and their association with risk perception of radiofrequency electromagnetic radiation (RF-EMR) exposure.
~\citet{Redmayne2017Phone} conducted an online questionnaire with fewer than $200$ women on their smartphone carrying preferences, revealing that $86\%$ carried their phones off-body, in the hand ($58\%$), a skirt/trouser pocket ($57\%$),and $15\%$ stored them against the breast (participants reported multiple placement). A complementary study by~\citet{zeleke2022mobile} investigated mobile phone carrying locations among $356$ men aged $18$–$72$, aiming to identify correlations between carrying positions and radio frequency electromagnetic radiation (RF-EMR) risk perception. These studies contribute to a broader understanding of gender, and even cultural-based differences in device placement and perceived health risks, but do not account for potential behavioural variations across different kinds of wearable technology, such as smartwatches, headphones, and tablets.

Wearable technology studies further examined smartwatch usage patterns, considering daily routines and influencing factors. A longitudinal analysis of 50 college students required participants to wear smartwatches for over $200$ days while tracking daily trends~\cite{jeong2017smartwatch}. Three smartwatch user groups were identified: \textit{work-hour wearers}, \textit{active-hour wearers}, and \textit{all-day wearers}~\cite{jeong2017smartwatch}, revealing context-driven smartwatch adoption, offering implications for usability optimization in wearable technology design. Generally, it has also been argued that proxemics (how people perceive and use space, and how that space impacts their communication and social interactions), weight distribution and user movement ability have to be taken into account for on-body wearables~\cite{Zeagler2017ISWC, Gemperle1998Wearability}, given the risks of wear and tear to the device (see~\cite{stein1998development}), user discomfort~\cite{sullivan2016designing}, and movement obstruction~\cite{tilley2001measure}. These concerns highlight the importance of appropriate wearable placement on the body.

% Since that time there were a few studies that did collect mobile data combined with self reports etc. for various purposes. One of them was ~\citet{busso2025diversityone}
Combining self-reports with sensor data has emerged as a powerful approach for understanding human activity, behaviour and context. Early datasets such as \textit{MDC} dataset ~\cite{laurila2013big} collected large-scale longitudinal smartphone data including coarse-grained sensors, mobility, and communication logs. Subsequent datasets have diversified in scope and modality: \textit{EgoADL} by~\citet{sun2024multimodal} uses audio, wireless, and motion sensors for scalable, privacy-preserving daily-life logging; \textit{SmartUnitn2}~\cite{li2022representing} integrates heterogeneous sensor streams with self-reports, questionnaires, and annotations to model situational context and personal habits; ~\citet{wilson2022qwantify}, recorded \textit{Qwantify} which captures in-the-moment smartphone-based experience sampling on desires alongside emotional and physical states. More recent multi-country and longitudinal datasets, such as \textit{DiversityOne} by~\citet{busso2025diversityone} or \textit{GLOBEM} by~\citet{xu2023globem}, further enable cross-country behaviour modelling, domain generalisation, and evaluation of behavioural models using self-reports and passive sensing data, including mental health labels. However, these datasets remain limited either in population coverage (geographically or socially wise) or generally do not account for device–user positioning, a critical factor affecting sensor signals and thus activity and context recognition algorithms. Moreover, due to collection date are outdated in terms of used technology and observed social interactions and habits.

A literature review by~\citet{yfantidou2023beyond} found that mobile and wearable computing
%UbiComp\footnote{UbiComp stands for Ubiquitous Computing. The design of wearables is a commonly researched topic in the UbiComp community given ubiquity of such devices.} 
research under-represents non-white and non-WEIRD\footnote{WEIRD population understood as population of Western, Educated, Industrialized, Rich, and Democratic countries~\cite{henrich2010weirdest}.} populations. Though, over time, research in wearable computing and Personal Informatics (PI)---the design and study of user interactions with tools like fitness trackers, smartwatches, smart rings, among others---is increasingly acknowledging that PI tools need to be more inclusive of under-represented individuals, such as Arabs and Muslims (e.g.~\cite{niess2021_IDontNeedAGoal, Ibrahim2024_TrackingDuringRamadan}), women (e.g.~\cite{Epstein2017MenstrualTracking, lu2024collaborative-pregnancy-tracking}), and disabled individuals (e.g.~\cite{carrington2015IDontTakeSteps}). Especially given that commercial PI tools tend to be designed with a `default' body in mind, there is a need to better understand how these tools can be more inclusive and accessible. However, when PI research aims to address this gap, it is mostly addressed from the perspective of metric design\footnote{With metrics, we refer to visualisations of personal data. Some examples of metrics that are typically visualised in PI tools include stress scores, heart rate, step counts, and sleep.} (see~\cite{Bentvelzen2023DerivedMetrics, Loerakker2025_LivEM}), and not necessarily from the perspective of how to make hardware design more appropriate and flexible for a variety of carrying habits.

Building on prior research, our study expands its scope to investigate a diverse range of wearable devices, including smartphones, smartwatches, and other wearables. We aim to investigate the actual carrying habits of participants recruited through a typical research channels-based experimental setting.

%To enhance inclusivity, we aim to diversify our participant pool in terms of culture, gender, education, and profession. 

\subsection{Past Guidelines on Hardware Design of Wearables} 
\label{sec:past_guidlines}

\begin{figure}[!t]
    \centering
    \includegraphics[width=\linewidth]{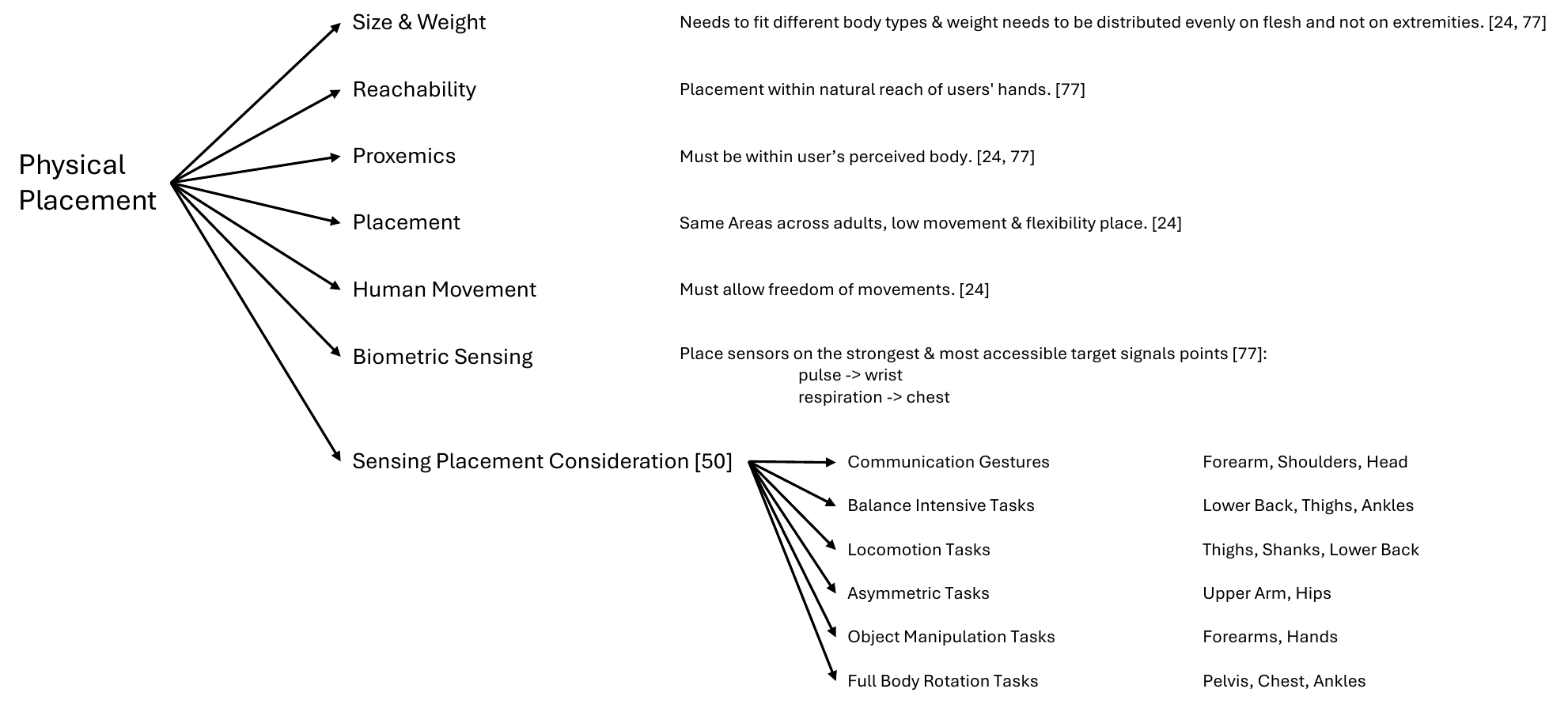}
    \caption{Synthesis of wearable hardware design guidelines from past research.}
    \label{fig:related_work}
\end{figure}

In the past a few articles were written with guidelines and tips for researchers and designers to follow while working on new devices.~\autoref{fig:related_work} presents a synthesis of relevant points and conclusions about device positioning derived from related work. 

The oldest was written by \citet{Gemperle1998Wearability} in 1998. Based on their extensive experience in device design in various fields the authors propose 13 guidelines for wearability including considerations on placement, adjustability, proxemics, sizing and more. They argue for human centric design across all types of bodies types and needs. However, the set of guidelines is quite broad and general, talking about comfort of placement or interaction without taking people's habits into consideration. Furthermore, due to its age, this work could not take modern technology into account. 

Building on the work of~\citet{Gemperle1998Wearability}, as well as extensive background research,~\citet{Zeagler2017ISWC} published a new set of guidelines that take the influence of following the previously proposed guidelines into consideration. Moreover, based on related work he assembled body maps that visually present possible sensor placements depending on the type of activity and context. The substantial research behind this work provides solid advice on what to consider during the design stage and presents examples of existing research where these guidelines were applied. Nevertheless, while the paper considers certain aspects of social acceptability (e.g. \textit{``In general, avoid touch-based interactions and displays within regions of the body associated with sexual activity or elimination of body waste.''}~\cite[p.~155]{Zeagler2017ISWC}), it does not examine the users' habits and patterns concerning the contexts in which wearable devices are actually worn. Which, as mentioned before, can diminish the amount of data collected or mislead the algorithm with data from different locations than those it was trained on, leading to reduced accuracy.. 

The first longitudinal study on smartwatch wearing behaviour was done by~\citet{jeong2017smartwatch} and included 50 participants. In the course of the experiments a few trends were discovered and used to derive user experience design implications. Captured \textit{in vivo}, the authors highlight issues that escaped during design of smartwatches, or emerged during their use (e.g. inconvenience of having different charger for smartwatch only). This article presents an important first step to understand people's habits and preferences. However, authors acknowledge some limitations like the usage of one specific type of smartwatch for their experiments or the limited number of study participants. 

\citet{ray2025w2w} presented a Where to Wear (W2W) system that is pointing to the optimal combination of the number of inertial measurement unit (IMU) sensors and their placement on the body for effective motion capture (MoCap) based on data simulation. They identified six categories (Communication Gestures, Balance Intensive Tasks, Locomotion tasks, Asymmetric Tasks, Object Manipulation Tasks, Full Body Rotation Tasks) of activities and provide guidelines for the best positioning for each one of them (\autoref{fig:related_work}). These guidelines are derived from locations that yield the best model performance when the simulated data are used as input. Although the authors account for movement noise, they do not consider deliberately “incorrect” data—such as data arising from improper sensor placement—that would simulate real-world user behaviour deviating from ideal usage patterns.

Overall, while existing guidelines are detailed and valuable for device design, their generalisability across countries, cultures, genders, and age groups remains uncertain. More critically, it is unclear how these guidelines apply to the populations typically recruited to test, evaluate, and validate such devices. This could lead to wrong data collection and thus invalid findings in the end

\section{Methodology}

Our inquiry was guided by two research questions: 
\begin{itemize}
    \item \textbf{RQ1:} \textit{How are on-body placements for wearable technologies influenced by user attributes?}
    \item \textbf{RQ2:} \textit{How do contextual factors influence users' on-body wearable placements?}
\end{itemize}

The first research question examines the influence of user demographics (e.g. age, gender, education) on the storage of wearable devices. The second question considers how contextual factors, such as time of day, activity, or environment, impact storage behaviour. Together, these questions investigate how context affects user behaviour.

To answer those questions, we conducted a demographically unrestricted, online questionnaire. Our study was approved by the ethics board of the primary researcher's affiliation.

In addition, the study is guided by a third research question:
\begin{itemize}
    \item \textbf{RQ3: } \textit{To what extent are assumptions about user characteristics, found in prior research, accurately represented?}
\end{itemize}

Through this research question, we examine which assumptions have been used in prior research, what types of guidelines exist, and how these relate to real-world data. To this end, we conducted a synthesis of related work (\autoref{sec:past_guidlines}) and compared these findings with the results of our survey (\autoref{sec:extending_old_design_guidlines}). In particular, we focused on identifying similarities, differences, and the extent to which assumptions are transferable to real-world contexts.

\subsection{Survey Content}

\begin{figure}
    \centering
    \includegraphics[width=\linewidth]{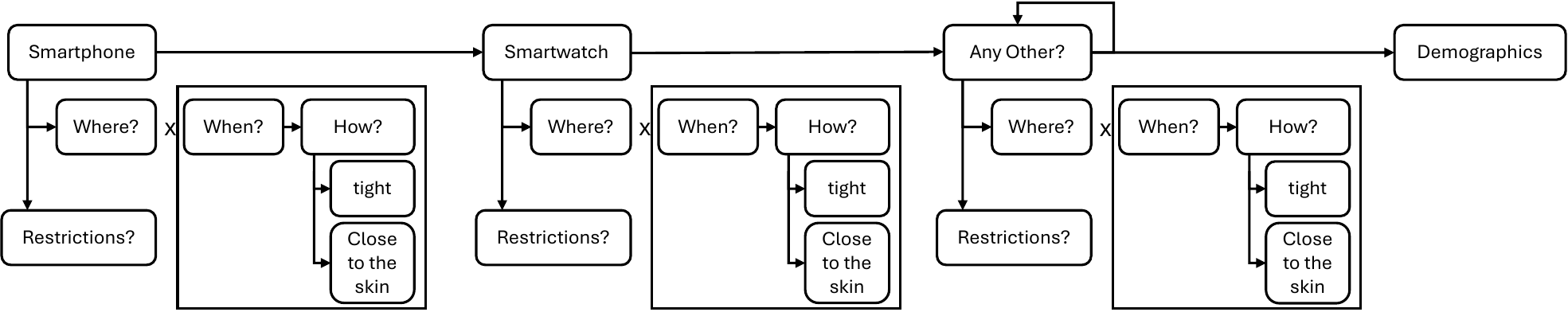} 
    \caption{Flowchart depicting the structure and contents of the questionnaire used in the study.}
    \label{fig:questionnaire_flowchart}
\end{figure}

%Our survey was offered in several languages [languages removed for anonymity] to increase accessibility and to ensure broad geographical coverage across countries. 
Our survey was offered in several languages (German, English, French, and Polish) to mimic the way studies are performed in international groups that are taking advantages from different cultures and countries.

The time to complete the survey ranged from 2 minutes 40 seconds to 27 minutes 19 seconds, with an average duration of 9 minutes 32 seconds. %In~\autoref{fig:questionnaire_flowchart}, a flowchart of the questionnaire structure is provided.
The survey was divided into four sections, as seen in~\autoref{fig:questionnaire_flowchart}: (1) smartphone usage; (2) smartwatch usage; (3) other wearable devices, and (4) demographic questions. The first and second sections of the questionnaire had the same structure, differing only in the type of device being addressed. For each device, participants were asked where they typically carry it. For each reported location, they were then asked during which times of the day or night (e.g. at home, outside, while sleeping) they use that placement, as well as how tightly the device fits and how close it sits to the skin. The third section followed the same structure but was designed to repeat automatically, allowing participants to report additional wearable devices as desired. In the forth section, demographic questions are asked.

\subsection{Participants}

% \begin{itemize}
%     \item (1) mirror the sample population that is present in related works on wearables/systems/prototypes (IMWUT, ISWC, MobileHCI) -> show what is true and what is not true (reseracher think how users use the devices, but they are using it differently) -> problematic cause wrong assumptions on wearable usage might lead to wrong implications/findings of research work
%     \item (2) possibly include a "counter example" (rwanda, low HDI) showcasing anomalies, or different pattersns -> this is just bonus; people use wearables differently in Africa (-> see CHI keynote, they went directly to the smartphone, we progressed over PC, laptop, ...)
% \end{itemize}

As described in the introduction, considering realistic wearable placement during the design phase is essential. Failure to do so may result in user discomfort, as well as misleading assumptions and, ultimately, flawed findings. Therefore, in this work, we examine whether the populations typically recruited for scientific experiments exhibit device-carrying habits that align with prior work and existing guidelines. If these guidelines fail to accurately represent even this commonly studied population, their applicability to the wider general public is likely to be even more limited.

In scientific research, participant recruitment is commonly conducted via mailing lists, advertisements in public spaces, and word of mouth; accordingly, we used these channels as well. Participants were compensated by being entered into a raffle with prizes totalling 200 euros.

% To reach a broad international audience, we used mailing lists, advertisements in public spaces, and word-of-mouth to recruit users of wearable technology. Participants were compensated by getting a chance to win prizes totalling 200 euros in a raffle.

\begin{figure}[!t]
  \centering
  \begin{minipage}{0.49\linewidth}
    \centering
    \includegraphics[width=\linewidth]{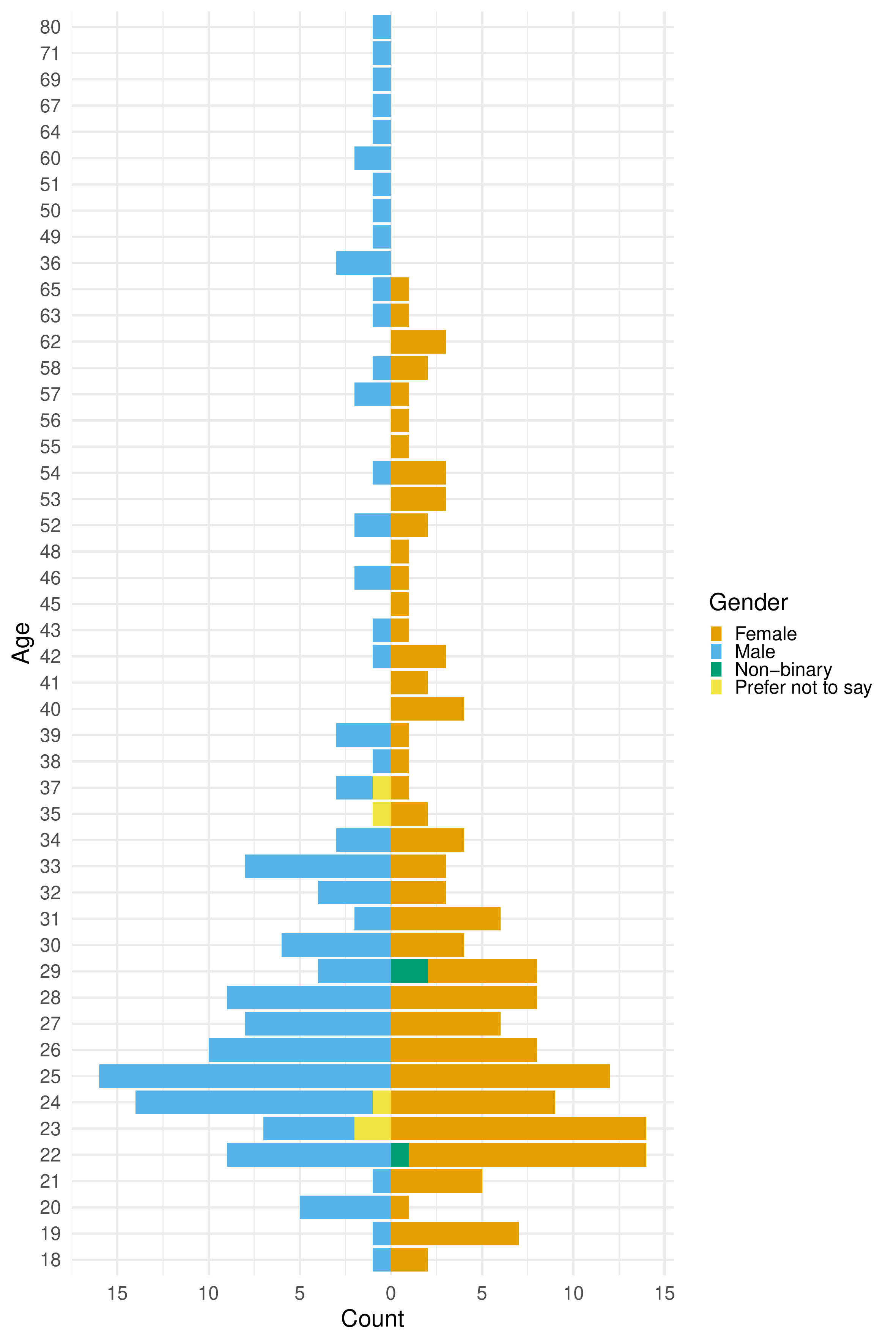}
    \caption{Distribution of age of participants grouped by gender.}
    \label{fig:age_vs_gender_hist}
  \end{minipage}
  \hfill
  \begin{minipage}{0.49\linewidth}
    \centering
    \includegraphics[width=\linewidth]{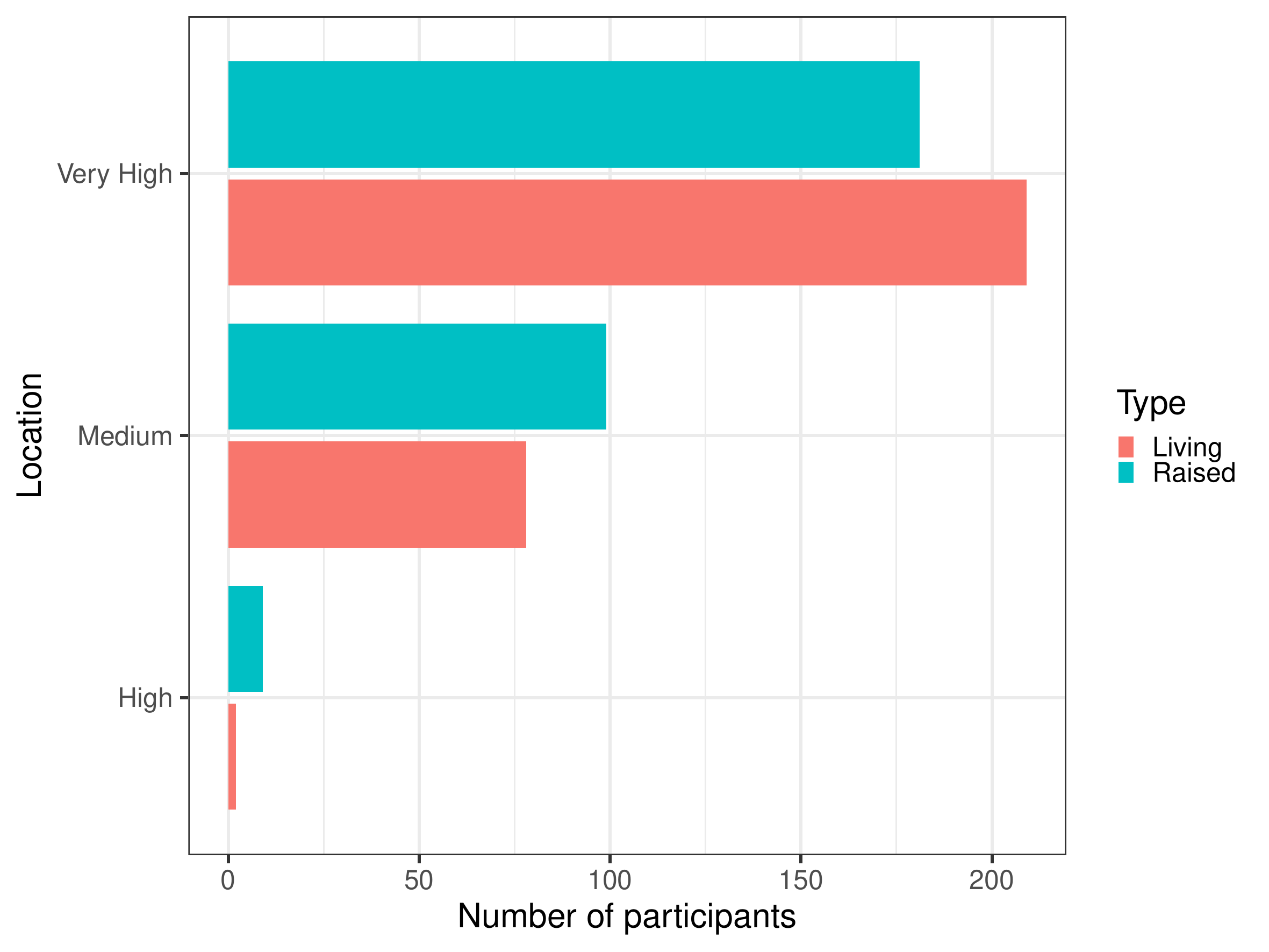}
    \caption{Histogram of the number of countries in which participants ever lived, grouped by country of raised and living. The countries are categorised according to the Human Development Index.}
    \label{fig:raised_living_graph_hdi}
  \end{minipage}
\end{figure}

%\autoref{fig:raised_living_graph} presents the distribution of all countries participants ever lived. We grouped them by places where participants grew up and where are currently living. Because of the way of distribution of the questionnaire every country except for Germany was primarily country of growing up.

In total, $n = 449$ participants were recorded, of which we included $n = 300$ in our final analysis. Reasons for exclusion included failing to complete the survey or to provide valid answers to mandatory fields. The gender distribution of our sample consisted of 150 females ($50\%$), 141 males ($47\%$), 3 non-binary ($1\%$), 1 other ($0.33\%$), and 5 who preferred not to share ($1.67\%$). The participants were between the ages of 18--80, with the age distribution skewed toward the younger generation ($M = 31.9$, $SD = 12.12$; see~\autoref{fig:age_vs_gender_hist}).

Besides basic demographic information, we also asked participants about their background, such as highest achieved education, handedness, country of residence, and occupation, to collect insights into whether contextual and environmental factors influenced their wearable use. 

In our survey, most participants were right-handed ($n = 268$), some left-handed ($n = 23$), and only a few ambidextrous ($n = 9$). Our sample had the following education levels: 21 ($7\%$) practical, 52 ($17.33\%$) high school, 96 ($32\%$) bachelor's, 104 ($34.67\%$) master's, and 27 ($9\%$) PhD degrees. The reported types of occupations were divided into the categories presented in~\autoref{tab:job_gender}.

\begin{table}[ht]
\centering
\caption{Number of participants by occupation type and gender; see \Cref{subsubsec:filter} for the inclusion criteria.}
    \begin{tabular}{lccc}
    \toprule
    \textbf{Occupation type} & \textbf{Total participants} & \textbf{Female participants} & \textbf{Male participants} \\
    \midrule
    Medicine \& Health      & 67 & 31 & 35 \\
    Computer Science         & 66 & 21 & 40 \\
    Humanities               & 33 & 27 & 6 \\
    Science                  & 33 & 15 & 18 \\
    Engineering              & 33 & 9  & 24 \\
    Business \& Economy      & 27 & 22 & 3 \\
    Education                & 14 & 9  & 5 \\
    Blue collar              & 11 & 4  & 7 \\
    Administration            & 12 & 10 & 1 \\
    Technicians              & 3  & 1  & 2 \\
    Architecture             & 1  & 1  & 0 \\
    \bottomrule
    \end{tabular}
\label{tab:job_gender}
\end{table}

Participants originated from various countries, including Germany, Poland, Rwanda, the United States, India, Egypt, France, Japan, Australia, China, Sweden, and Indonesia (the distribution among countries has been removed for review anonymization). 

Among participants of the survey, 153 (51.00\%) reported living in the city, 135 (45.00\%) in the countryside, and 7 (2.33\%) in towns. Additionally, 5 (1.67\%) participants reported living in a few types of settlements at the same time, likely reflecting the travelling nature of their work.

In our sample, 289 participants (96\%) used smartphones, 92 (31\%) smartwatches, 83 (28\%) earphones, 18 (6\%) headphones, 11 (4\%) location trackers, 4 (1\%) cardiac monitors, 2 (1\%) tablets, 2 (1\%) smart rings, 2 (1\%) smart glasses, and 1 (0.3\%) Coros pod. It is important to note that some participants reported using more than one device at a time.

\subsection{Analysis}
To identify associations between demographic characteristics of our sample and wearable device usage patterns, we employed an automated exploratory data analysis approach. Given the large number of variables included in our survey, a systematic evaluation of all possible pairwise combinations was conducted. It is important to note that all percentage data reported in this paper were calculated by dividing the number of occurrences of interest by the total number of participants.

\subsubsection{Variable Selection}
Our independent and dependent variables (IVs, DVs) were selected from `Device', `Age', `Gender', `Education Level', `Work Status', `Country of Origin', `Country or Residence', `Parental Education Group', `Friend Group Influence', `Environment Score', `Place' of wear, `Time' of use, `Tightness' preference, and `Clothes' worn during usage.

\subsubsection{Combination Generation}
All unique combinations of one, two and three IVs, and DVs were generated. This resulted in a total of 5655 distinct combinations of combinations for analysis (assuming $C^i_I$ is number of combinations of independent variables, and $C^i_D$ is a number of combinations of dependent variables: $Sets_1 = C^1_I*C^1_D$, $Sets_2 = C^2_I*C^1_D$, $Sets_3 = C^3_I*C^1_D$ which in total gives: $Sets = C^1_I*C^1_D + C^2_I*C^1_D + C^2_I*C^1_D$).
%combination(13,1) = 13, combination(13,2) = 78, combination(13,2) = 286).
Combinations with more than three IVs were also evaluated but did not yield any statistical significance.

\subsubsection{Data Preprocessing \& Filtering}
\label{subsubsec:filter}
Prior to testing, categorical variables were consolidated where appropriate to improve statistical power and interpretability. For several independent variables (IVs), we filtered out conditions that lacked sufficient responses to ensure the generalisability of our findings. For instance, for the \textit{Device} IV, we first combined the categories \textit{earphones} and \textit{headphones} due to their similar purpose and usage. We then included only \textit{smartphones}, \textit{smartwatches}, and \textit{earphones} in our analysis, as these were the only devices representing more than 10\% of all responses.

For the \textit{Gender} IV, only data from participants identifying as male or female were included, as the number of other reported answers were too low to draw any meaningful conclusions.

To analyse the influence of country while maintaining the full dataset, we used the Human Development Index 2025 (HDI) \cite{UNDP_HDI}. The HDI ranges from 0 to 1 and categorizes countries based on multiple dimensions, such as `long and healthy life', `knowledge', and `decent standard of living'). It is updated each year by the World Health Organization (WHO). Depending on the HDI value, countries are divided into four categories: Very High (>0.800), High (0.700 - 0.799), Medium (0.550 - 0.699), and Low (<0.550). Most participants in our survey reported having grown up in Very High HDI countries ($n=187$, $62.8\%$), followed by Medium HDI countries ($n=102$, $34.2\%$), with only a few raised in High HDI countries ($n=9$, $3.2\%$). 

To obtain a more complete picture, we also analysed how many participants currently live in countries within those categories. The general trend remains similar, although 30 participants moved to countries in a higher HDI category and one person moved to a country in a lower HDI category (see~\autoref{fig:raised_living_graph_hdi} for the distributions; see appendix for exact numbers per country).

During the analysis of time-related wearing patterns, we examined only the categories: \textit{Active time}, \textit{During work hours}, \textit{Sport}, \textit{While in bed}, and \textit{Outside}, as other times were not reported frequently enough to yield statistically meaningful results when dividing the data by different demographic variables. Some of those times were congregated from other responses of participant, thus: 

\begin{itemize}
    \item \textbf{Active time:}  was created by combining two categories: \textit{While not in bed} and \textit{During the day} as those both describe a time that is spend actively doing tasks ranging from leisure, through house chores to work
    \item \textbf{Outside:}  was combined from \textit{on the way}, \textit{outside}, \textit{travelling}, \textit{using public transport} as all of them combine notion of travelling in vehicles or being outside.
\end{itemize}

\subsubsection{Statistical Analysis}
The association between each IV and DV pair was assessed using a Chi-Square Test of Independence ($\chi^2$) \cite{pearson1895x}. Contingency tables were constructed to summarize observed frequencies across all combinations of categorical values. 
To ensure the validity of the test results, we evaluated expected cell frequencies within each contingency table. Cells with an expected frequency less than 5 were considered potentially problematic. If more than 10\% of cells had an expected frequency below this threshold (min. expected threshold = 5), the corresponding chi-squared test result was excluded from further consideration.
A p-value threshold of 0.001 was applied to identify statistically significant associations.

\section{Results} \label{sec:findings}
\input{tab_finding}

This section presents the findings from our study, structured into two main parts. First, we report the results of our statistical tests and observations, as summarized in~\autoref{tab:statistical_results}, focusing on how user demographics—such as age, education, work background, country of origin and residence, and social influences (e.g. parents, friends, environment)---affect wearable device usage, on-body placement, and usage time. Notably, gender was found to influence only the placement of devices. Second, we explore how these demographic factors interact with time-of-day and context to shape placement behaviour across daily routines.

%       Device Gender   n
% 1  earphones Female  54
% 2  earphones   Male  42
% 3 smartphone Female 148
% 4 smartphone   Male 133
% 5 smartwatch Female  51
% 6 smartwatch   Male  40

\subsection{Demographic Influence}

\begin{figure}
    \centering
    \includegraphics[width=\linewidth]{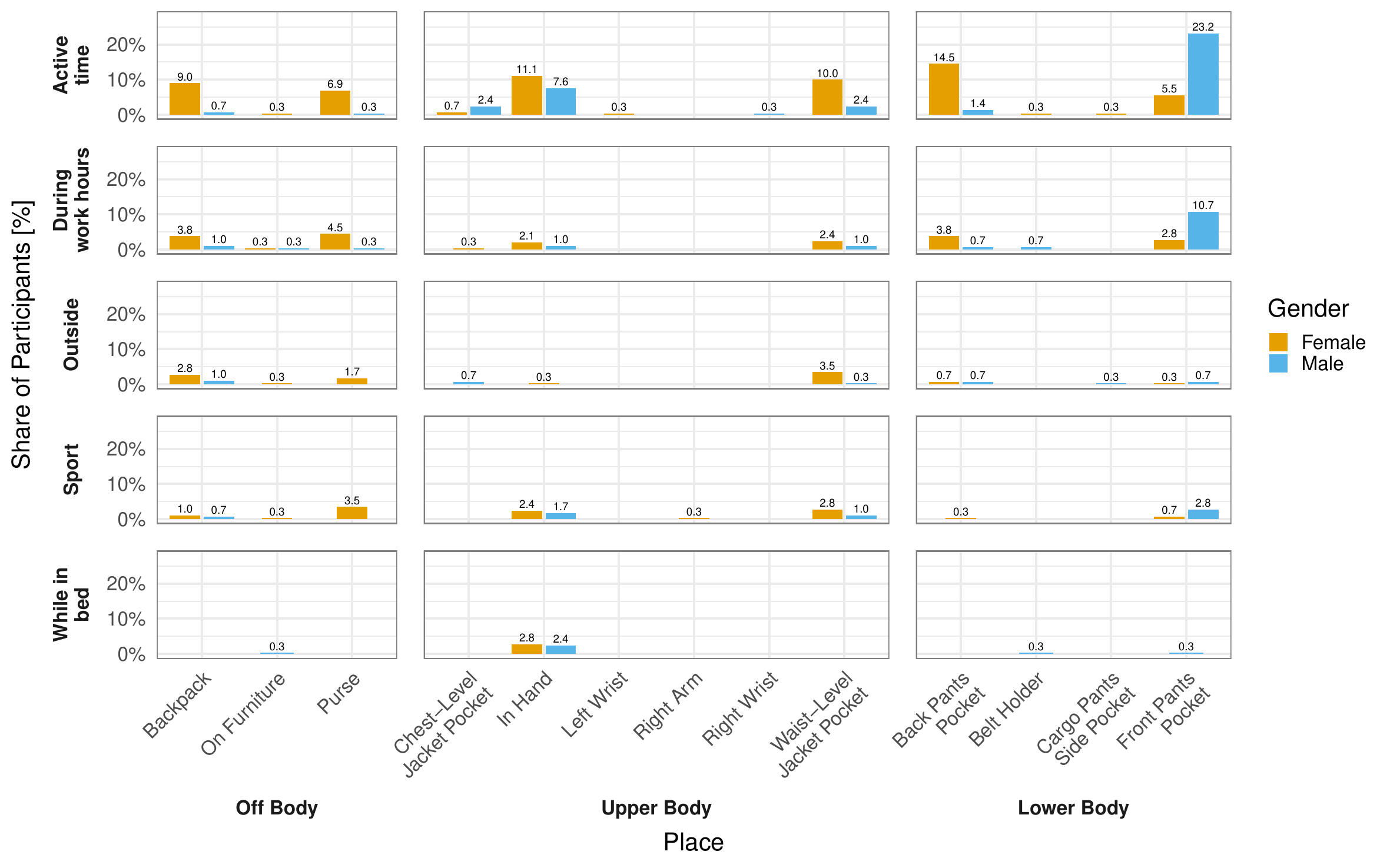}
    \caption{Bar plot showing the distribution of smartphone carrying locations by time of day and gender; see \Cref{subsubsec:filter} for the inclusion criteria. For each time period, percentages were calculated separately as the number of responses for each location category divided by the total number of study participants.}
\label{fig:place_by_device_by_times_reduced_by_gender_other_smartphone_hit}
\end{figure}

\begin{figure}
    \centering
    \includegraphics[width=1\linewidth]{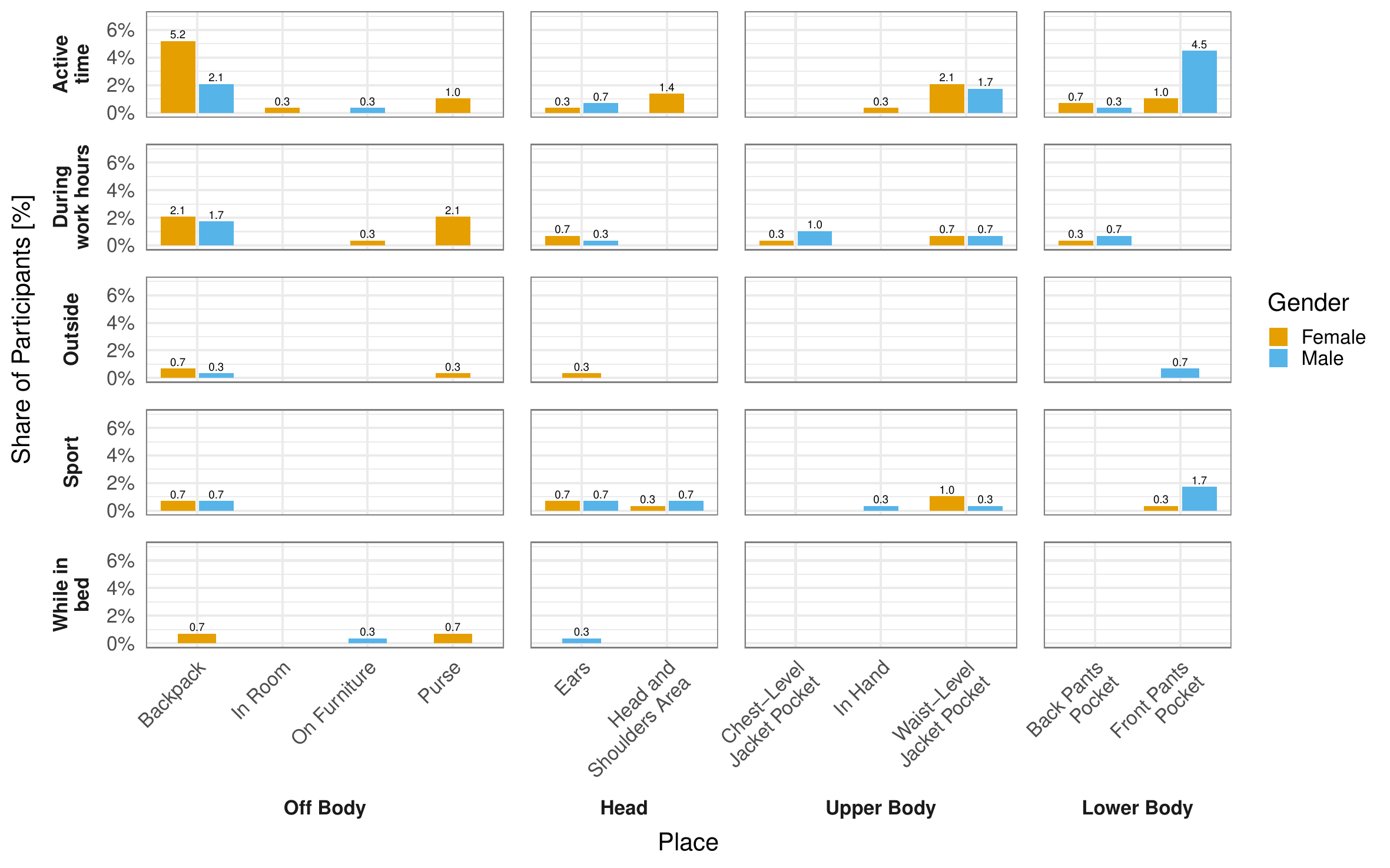}
    \caption{Bar plot showing the distribution of earphones carrying locations by time of day and gender; see \Cref{subsubsec:filter} for the inclusion criteria. For each time period, percentages were calculated separately as the number of responses for each location category divided by the total number of study participants.}
    \label{fig:place_by_device_by_times_reduced_by_gender_other_earphones_hit}
\end{figure}

\subsubsection{Age.} 
Across all age groups, smartphone was the most commonly used wearable device. However, age-specific placement patterns emerged. Younger participants (under 28 years) showed a stronger preference for earphones and were more likely to carry their smartphones in hand. Older individuals—especially women—were more likely to store their smartphones in purses or bags. Smartwatches were typically worn on the left wrist regardless of age, although younger users showed more placement variability.

\subsubsection{Gender.} 
\label{subsubsec:results_gender}
Gender significantly influenced both device usage and placement. For smartphones, men primarily used front pants pockets throughout the day ($33.9\%$ men in opposition to $5.37\%$ women), while women preferred back pants pockets or waist-level jacket pockets ($14.1\%$ and $9.73\%$, respectively). During work hours and physical activity, women tended to store their phones in purses ($4.36\%$) or backpacks ($3.69\%$), whereas men continued to use front pants pockets ($16.1\%$) or held the phone in hand. At night, both genders reported holding the phone in hand, likely indicating active use in bed rather than passive storage.

Earphone use patterns were similar across genders during the day and while exercising. However, storage preferences differed: men favoured pants pockets, while women preferred purses and waist-level jacket pockets. Both used backpacks for storage. Smartwatch use showed comparable trends, most participants wore them on left wrist ($28.6\%$ of women, $41.9\%$ of men), particularly during active or sport periods, with few choosing alternative placements.

\subsubsection{Education.}
% Educational background showed no strong direct influence on device placement, but was correlated with age and occupational factors. Participants with higher education—especially Master's and PhD holders—demonstrated greater variability in placement. PhD participants primarily stored earphones in backpacks or front pants pockets. Master’s graduates reported active earphone usage throughout the day, with backpacks and front pants pockets being common storage choices. High school graduates reported usage primarily during sports, often storing headphones on the head or shoulders. Bachelor’s graduates also favoured backpacks but preferred waist-level jacket pockets during sports.

Educational background showed limited direct influence on device placement but was associated with occupation and age. During active time, smartphone storage was relatively consistent: front pants pockets were common among Bachelor’s, Master’s, and PhD holders ($20.1\%, 17.5\%$, and $17.1\%$, respectively), while high school graduates more often used back pants pockets ($16\%$) or held the phone in hand ($15\%$).
Earphone storage patterns were more variable, during work and sport, most users kept them around the ears, whereas during active time they were typically stored elsewhere. 
Smartwatches were predominantly worn on the left wrist across all education levels, particularly among Master’s ($40\%$) and Bachelor’s graduates ($29.4\%$) during active periods.

\subsubsection{Occupation.}
\label{subsubsec:occupation}
% Participants from the medical and health fields most frequently stored smartphones in front pants pockets during work and active periods. Computer scientists also favoured this location but shifted toward waist-level jacket pockets during active times. They also reported frequent smartphone use in bed. Humanities and business participants displayed more variable patterns with no dominant placement. Engineers and STEM professionals generally favoured front pants pockets; engineers also used hand-held placements during active times, while scientists preferred back pants pockets or jacket pockets. Notably, scientists often reported placing smartphones on furniture, suggesting greater awareness of device handling. Medical professionals generally reported major tendencies in storing their smartphones in all kinds of pockets: front pants pocket, back pants pocket, waist level jacket pockets etc. For earphones, health professionals preferred front pants pockets, followed by backpacks and waist-level jacket pockets. Computer scientists stored earphones in pants pockets and backpacks and reported active in-ear usage throughout the day, except outdoors. Humanities and business participants primarily used backpacks and jacket pockets, with business professionals also using purses. Scientists and engineers showed no strong patterns but leaned toward backpack storage.

Smartphone placement varied by occupation and time. Medical and health professionals mainly used front pants pockets during work ($42.9\%$) and active periods ($34.4\%$). Computer scientists preferred front pants pockets during work ($36.4\%$) and active time ($34.9\%$), with in-hand use ($25.4\%$). Engineers primarily used front pants pockets during working hours ($46.2\%$) and active time ($31.2\%$), also holding devices in hand ($28.1\%$). Scientists distributed smartphones between front ($33.3\%$) and back pants pockets ($24.2\%$). Humanities and business participants were more variable, using backpacks, jacket pockets, and purses depending on activity.
Earphone placement followed similar patterns. Health professionals used front pants pockets ($36.4\%$), backpacks ($27.3\%$), and jacket pockets ($27.3\%$) during active times. Computer scientists preferred front pants pockets ($33.3\%$) and backpacks ($28.6\%$), with frequent in-ear use during working hours. Humanities and business participants favoured backpacks ($42.9$–$55.6\%$) and purses ($33$–$50\%$), while scientists used backpacks ($33.3\%$) and jacket pockets ($16.7\%$).
Smartwatches were mostly worn on the left wrist, especially during active time (computer science $85.7\%$, engineers and health $80\%$) and sport (health $83.3\%$, humanities $100\%$), with consistent left-wrist use across occupations during work and in-bed periods.

\subsubsection{HDI and Country of Origin.}
% Participants from very high HDI countries generally owned more devices ($Mean_{\text{Very high HDI}} = 1.92$ vs $Mean_{\text{Medium HDI}} = 1.57$) and showed more diverse placement behaviours. While front pants pockets remained the most common smartphone location, other placements such as back pockets, jacket pockets, and in-hand use were also frequently reported. Medium HDI participants showed more uniform behaviour, typically owning a single device and using the front pants pocket for storage.

Participants from very high HDI countries generally owned more devices as shown by $Mean_{\text{Very high HDI}} = 1.92$ vs $Mean_{\text{Medium HDI}} = 1.57$ and exhibited more diverse placement behaviours. During active time, front pants pockets were common ($24.5\%$), but back pockets ($20\%$), jacket pockets ($16.5\%$), in-hand use ($15.5\%$), and backpacks ($11.5\%$) were also reported with smartphone. Medium HDI participants showed more uniform behaviour, primarily using front pants pockets ($42.9\%$) and in-hand ($28.6\%$) storage.  
Smartwatches were predominantly worn on the left wrist, especially during active time (very high HDI $92.9\%$, medium HDI $69.2\%$) and sport (very high HDI $93.8\%$, medium HDI $75\%$), with consistent left-wrist use during work and in-bed periods.  
Earphone placement followed similar patterns. Very high HDI participants used backpacks ($34.8\%$), front pants pockets ($21.7\%$), and jacket pockets ($19.6\%$) during active time, while medium HDI participants favoured front pants pockets ($35.7\%$) and backpacks ($28.6\%$). Hybrid participants used varied locations during work and sport, including backpacks, purses, and ears (up to $28.6\%$), reflecting flexible handling habits.

Participants raised in medium HDI countries but currently living in very high HDI contexts exhibited hybrid behaviour. Their preferences still leaned toward front pants pockets but included more varied placement during specific activities such as work or sports. Those who moved to higher HDI countries more commonly owned three devices, particularly if they held a Master’s degree or worked in computer science.

\subsubsection{Device Ownership.}
Most participants reported owning two wearable devices. This trend was most evident among individuals living in cities, with higher education levels, or in very high HDI countries. Participants in technical and humanities fields often owned multiple devices. In contrast, participants from medium HDI countries or those with bachelor’s degrees typically owned only one device. Among participants who moved to higher HDI countries, those with a Master’s degree or a background in computer science were more likely to own three devices. Women in this group often reported owning three or four devices, while men leaned toward two or three.

% Previously in Time centered section (next)

% \paragraph{Time of Day and Activity Context.}
% Smartphone placement changed significantly depending on time of day and activity. During active hours, men consistently used front pants pockets, while women transitioned to backpacks or purses. At night, most participants—regardless of gender—held smartphones in hand, indicating likely use before sleep. During physical activity, smartphones were commonly carried in hand or placed in jacket pockets. Women were more likely to store them in bags. These behaviours suggest participants may have reported break-time usage, as using phones during active exercise may not be practical.

% Smartwatch placement was relatively consistent throughout the day. Most participants wore the device on the left wrist, likely due to right-handed dominance.

% Earphones were used most actively during sports and daytime activities. When not in use, they were stored in backpacks by both genders. However, men more often used front pants pockets, while women preferred jacket pockets or purses. Some participants stored earphones on the head or shoulders, likely indicating headphone use.

\subsubsection{Urban vs. Rural Differences.}
% Urban and rural participants shared common primary smartphone placements, particularly the front pants pocket. However, urban users more frequently held phones in hand, potentially due to security, privacy, or active use. Rural participants leaned more toward backpacks and purses for secondary storage.
%For earphones, urban residents reported usage throughout the day and work hours, with backpacks as the primary storage location. Rural participants used earphones mainly during sports, again preferring backpacks for storage regardless of activity.
Urban and rural participants shared similar primary smartphone placements, with front pants pockets being the most common during active time (City: $22.4\%$, Country Side: $38.1\%$) and work hours (City: $33.9\%$, Country Side: $41.9\%$). However, urban users more frequently held phones in hand during active periods (City: $24.3\%$) and while in bed (City: $87.5\%$), suggesting greater mobile use for security, privacy, or convenience. Rural participants preferred backpacks and purses for secondary storage, particularly during sports (Backpack: $13.6\%$, Purse: $9.1\%$) and outside activities (Backpack: $20\%$, Purse: $20\%$).  
Smartwatch placement was largely consistent, with the left wrist predominant in both urban (Active: $87.5\%$, Sport: $86.7\%$) and rural participants (Active: $72.7\%$, Sport: $100\%$). Right wrist and waist-level pockets were minor alternatives.  
Earphone usage reflected activity and environment. Urban participants used earphones throughout the day and during work, storing them in backpacks (Active: $25\%$, Work: $35.3\%$) or front pants pockets (Active: $25\%$, Work: $11.8\%$). Rural participants mainly used earphones during sports (Backpack: $14.3\%$, Ears: $28.6\%$) and outdoor activities, preferring backpacks and front pants pockets. In-bed earphone use was similar across locations, often stored in backpacks or in-hand.

\subsection{Contextual Influences Across the Day}
This section explores how on-body device placement varies throughout the day and across activity contexts, highlighting how these patterns intersect with user demographics. These insights inform the analysis of real-world sensor placement and will support guidelines for future HAR implementations.

\begin{figure}
    \centering
    \includegraphics[width=\linewidth]{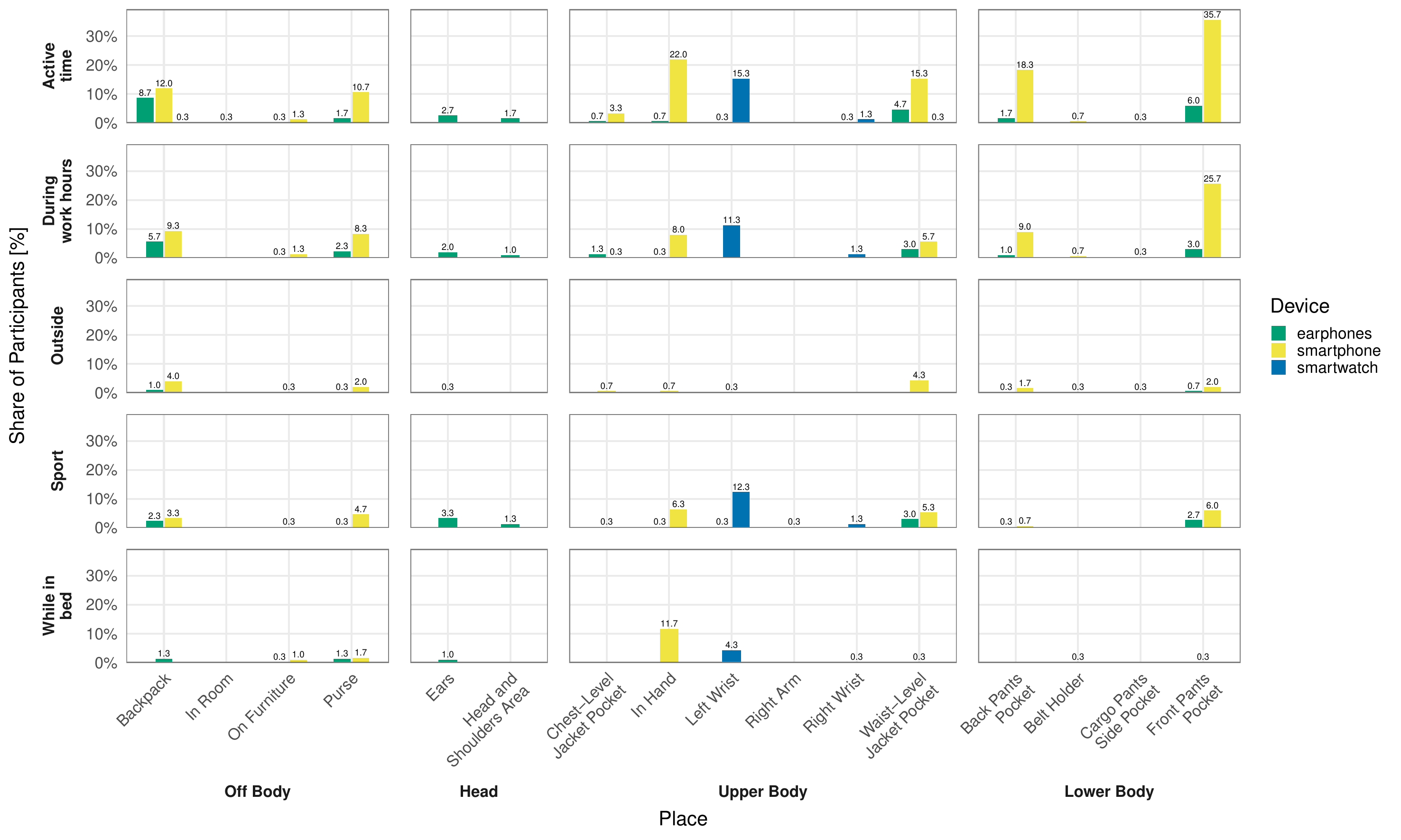}
    \caption{Bar plot showing the distribution of device carrying locations by time of day. Percentages were calculated separately for each time period as the number of responses for each location category divided by the total number of study participants.}
    \label{fig:placeholder}
\end{figure}

The most critical factor influencing where a device is kept is the activity context, which defines a clear functional distinction between active use locations (primarily the hand and ears) and passive storage locations (such as pockets, bags, and furniture). This analysis breaks down observed habits across five distinct periods, revealing strong temporal dependencies in device placement.

\subsubsection{Active Time Habits}
% During general \textit{Active Time}, users prioritize secure, high-accessibility storage. The Smartphone exhibits its highest placement frequency across the entire dataset in the front pocket of your pants, confirming this location as the default, primary carrying position for mobility. This positioning ensures both security and easy access. Earphones are primarily in storage during this period, showing high counts in the hand, Back pants pocket, and the purse. Usage is intermittent rather than sustained. As expected, the Smartwatch placement is stable, remaining consistently on the left wrist or right wrist, reflecting its continuous-wear function.
During general \textit{Active Time}, users prioritize accessibility and secure storage. Smartphones are most frequently carried in the front pants pocket ($29.6\%$), followed by in-hand ($18.3\%$), waist-level jacket pockets ($12.7\%$), backpacks ($9.97\%$), and purses ($8.86\%$), reflecting a balance between mobility and quick access. Earphones are commonly stored in backpacks ($29.9\%$), front pants pockets ($20.7\%$), and waist-level jacket pockets ($16.1\%$), with lower proportions in the ears or purses, indicating intermittent use rather than continuous engagement. Smartwatches remain predominantly on the left wrist ($86.8\%$), confirming consistent wear and continuous functionality during active periods.

\subsubsection{During Work Hours Habits}
% The patterns observed During work hours suggest a need for both secure storage and rapid intermittent access. The Smartphone is still heavily favored in the front pocket of your pants but also shows a high presence in the hand, implying frequent desk-side checks or quick use, potentially for focused professional communication. Earphones are predominantly found in storage locations, notably the front pocket of your pants and the purse, indicating that active in-ear use is minimal and the devices are being securely stowed while the user focuses on work tasks. The Smartwatch maintains its standard placement on the wrist.
During work hours, device placement reflects a balance between secure storage and rapid intermittent access. Smartphones are most frequently carried in the front pants pocket ($37.4\%$), followed by backpacks ($13.6\%$), back pants pockets ($13.1\%$), in-hand ($11.7\%$), purses ($12.1\%$), and waist-level jacket pockets ($8.25\%$), suggesting frequent desk-side checks and quick use for professional communication. Earphones are primarily stored rather than worn, with notable placements in backpacks ($28.3\%$), front pants pockets ($15\%$), waist-level jacket pockets ($15\%$), purses ($11.7\%$), and occasional in-ear use ($10\%$), indicating intermittent engagement. Smartwatches remain predominantly on the left wrist ($89.5\%$), consistent with continuous monitoring and accessibility during work.

\subsubsection{Outside Activity Habits}
%When users are Outside, device placement reflects a balance between security during transit and active functional use. Earphones placement shows a significant spike in the ears, indicating intentional, sustained listening for media consumption or calls while moving. The Smartphone storage splits between the Back pants pocket and the hand. This split suggests users either opt for secure but accessible pocket storage or hold the device for active use, likely for navigation or communication. The Smartwatch placement remains stable on the wrist.
During outside activities, device placement balances security and active use. Smartphones are primarily stored in waist-level jacket pockets ($26.5\%$) and backpacks ($24.5\%$), with smaller proportions in the front pants pocket ($12.2\%$), purses ($12.2\%$), and back pants pockets ($10.2\%$), suggesting users choose either secure, accessible storage or hold the device for active interaction such as navigation or calls. Earphones show a notable presence in the ears ($23.8\%$), indicating intentional media consumption or communication while moving. The smartwatch remains consistently on the left wrist ($90.2\%$), supporting continuous monitoring and accessibility during mobility.

\subsubsection{Sport and Exercise Habits}
%The Sport context is characterized by the need for active, immediate utility. All three devices (smartphone, smartwatch and earphones) show high usage in the hand, suggesting interaction is required for tracking metrics, changing music, or viewing timers. Earphones show one of their highest counts in the ears, confirming them as necessary accessories for exercise media. The Smartphone is also frequently in the front pocket of the pants or a purse, but its high placement in the hand indicates it is often retrieved for active control. The Smartwatch remains on the wrist, serving its primary function as an exercise tracker.
In the Sport context, devices are used for immediate, active interaction. Smartphones are most commonly in the front pants pocket (22.0\%) or held in hand (23.2\%), with additional use in purses (17.1\%) and waist-level jacket pockets (19.5\%), indicating frequent retrieval for exercise management. Earphones show high placement in the ears (23.8\%) and on the waist-level jacket pocket (21.4\%), confirming their role in exercise media, while some users store them in backpacks (16.7\%) or front pants pockets (19.0\%). Smartwatches remain primarily on the left wrist (90.2\%), consistently serving as exercise trackers. Overall, device placement reflects a balance between accessibility and active monitoring during physical activity.

\subsubsection{While in Bed Habits}
% The context \textit{While in bed} marks a major shift from portable storage to a pattern of stationary or final-use engagement. The Smartphone placement changes dramatically, shifting from pants pockets to stationary storage on furniture (like a night stand) or being held in the hand for final evening use. Earphones still show a high count in the ears, suggesting sustained listening (such as podcasts or sleep soundscapes). The Smartwatch remains on the wrist, strongly suggesting its utilization for sleep tracking rather than active timekeeping.
During the \textit{While in bed} context, smartphones shift from pockets to being held in hand (76.1\%) or placed nearby in purses (10.9\%), indicating final-use activities before sleep. Earphones remain in the ears, supporting continuous listening for podcasts, audiobooks, or sleep soundscapes. Smartwatches stay on the left wrist (92.9\%), emphasizing passive monitoring like sleep tracking rather than active use. Overall, this context shows a clear transition from mobile to stationary device interaction.

\section{Discussion} \label{sec:discussion}
A comprehensive interpretation and application of our findings require an understanding of the underlying factors that shaped the reported habits. Accordingly, we first analyse the influence of cultural trajectories, occupational typologies, and associated demographic determinants on wearable device usage. This examination elucidates the socio-technical parameters governing the evolving landscape of device usability and informs projections of future adoption patterns. Subsequently, we integrate our empirical insights with established design guidelines, refining them in light of contemporary trends to promote enhanced adaptability, relevance, and user-centred applicability.
Throughout this section, it is important to note that our sample represents a population commonly recruited in related research contexts. Although it is not intended to be fully generalisable across all cultures and social groups, it offers an indication of the direction in which many trends may evolve.

\subsection{Contextualizing Results in Culture and Society}

The distribution of wearable and device placement patterns across users is not random but shaped by social, occupational, and environmental context. Factors such as gendered clothing design, professional requirements, and living environments influence both the accessibility of body areas and the functional logic behind where devices are stored or worn. Understanding these contextual dimensions helps explain why certain algorithmically ideal placements are under-represented in real-world scenarios, while others—though less optimal in theory—dominate user practice.

\paragraph{Gender Differences}
Gender plays a significant role in shaping available storage options and, consequently, habitual device placements. As highlighted by~\citet{DiehmThomas2018} and~\citet{townsend2023pockets}, men’s trousers typically feature pockets that are not only more prevalent but also nearly twice as large as those found in women’s pants. This structural clothing disparity means that women often resort to alternative storage solutions, such as purses, handbags, or jackets, which naturally diversifies the number and distribution of device locations as visible on~\autoref{subsubsec:results_gender}. As a result, women’s placement data reflects a broader spatial variety, encompassing both body-proximal (e.g. jacket pockets) and external storage areas (e.g. backpacks, handbags). On the heat map (\autoref{fig:place_by_device_by_times_reduced_by_gender_other_smartphone_hit}) it is characterized by the absence of a bright spot on the female side, instead depicting a more uniform distribution. In contrast, men’s device usage is heavily concentrated in the front pants pockets, which serve as the primary, secure, and easily accessible location. This asymmetry explains why, even though women exhibit a wider range of placement contexts, the front pants pocket remains the single most frequently used location overall—driven by the homogeneity of male usage \autoref{fig:place_by_device_by_times_reduced_by_gender_other_smartphone_hit}{}.

\paragraph{Occupation and Professional Context}
Occupational factors further modulate device wearing patterns through functional and normative constraints. Certain professions, particularly those requiring continuous mobility or rapid accessibility (e.g. medicine, construction, or technical trades), favour placements that balance security with immediate reach---most notably the front pants pocket (as reported by more than $33\%$ participants in those professions in~\autoref{subsubsec:occupation}). These occupations often feature standardized uniforms with relatively gender-neutral pocket structures, reducing clothing-based disparities. In contrast, roles involving stationary or cognitively intensive work (e.g. scientists, engineers, researchers) tend to exhibit higher variability in placement, including frequent use of desks, furniture surfaces, or handheld use during note-taking or device-based referencing (more than $28\%$ of engineers reported holding their smartphone in hand during work hours).

In case of professions that need to be able to be contacted frequently and with ease in urgent cases, there is a visible trend of storing devices in places that are secure enough to stay there in during sudden movements but easy enough to get out for quick communication purposes. On of the representatives of that type of occupation is medicine in which the most trending type of places were different pockets, which are big and gender neutral (\autoref{subsubsec:occupation}), in scrubs. Furthermore, auditory device use (e.g. earphones) also follows occupational logic—while some knowledge-based jobs (such as in computer science or design) allow or even require earphones for virtual meetings or background media, others that demand full concentration (like medical or manual fields) naturally suppress such usage. Thus, the occupational environment not only defines the physical constraints of placement but also shapes behavioural norms of engagement with wearable and mobile technology.

\paragraph{Living Environment}
Finally, lifestyle differences linked to living environments, particularly between rural and urban settings, introduce further variation in device placement and usage intensity. As demonstrated in studies by~\citet{fan2014rural},~\citet{christiana2021active}, and~\citet{pelletier2023work}, rural residents typically engage more frequently in household and maintenance-oriented physical activities, while urban dwellers tend toward shorter, more structured, and higher-intensity activity bouts, such as gym workouts or commuting. These lifestyle differences affect both accessibility needs and preferred storage methods. In rural contexts, devices are more often stored in locations that minimize interference with manual or domestic tasks, such as backpacks, purses, or nearby furniture surfaces. In urban settings, by contrast, users more frequently maintain direct interaction with their devices—holding them in hand during transit or activity transitions ($24.3\%$ of city dwellers are keeping phone in hand while only $11.9\%$ of rural residents reported this habit). This reflects not only behavioural convenience but also the increased integration of the smartphone into social and navigational routines typical of dense urban environments.

\paragraph{Summary}
In summary, the interplay between gendered clothing design, occupational function, and lifestyle environment collectively explains much of the observed divergence between theoretically optimal device placements and actual user habits. While algorithmic models often assume consistent body-coupled sensor positions, real-world practice reflects a complex negotiation between anatomical comfort, accessibility, social acceptability, and the physical affordances of clothing and context. These findings highlight the need for context-sensitive adaptation in wearable sensing algorithms and user experience design.

\subsection{Extending Old Design Guidelines} \label{sec:extending_old_design_guidlines} % Previous Guidelines vs our findings

While our findings reflect the real-world practices and habits of wearable users, prior guidelines predominantly approach wearable placement from an engineering-oriented perspective (e.g. sensors positioned on the ankle), grounded in anatomical, physiological, and algorithmic considerations. To meaningfully compare our empirical insights with these theoretical recommendations and answer \textbf{RQ3}, a common analytical frame was required to bridge practice and theory.

\begin{figure}
    \centering
    \includegraphics[width=0.95\linewidth]{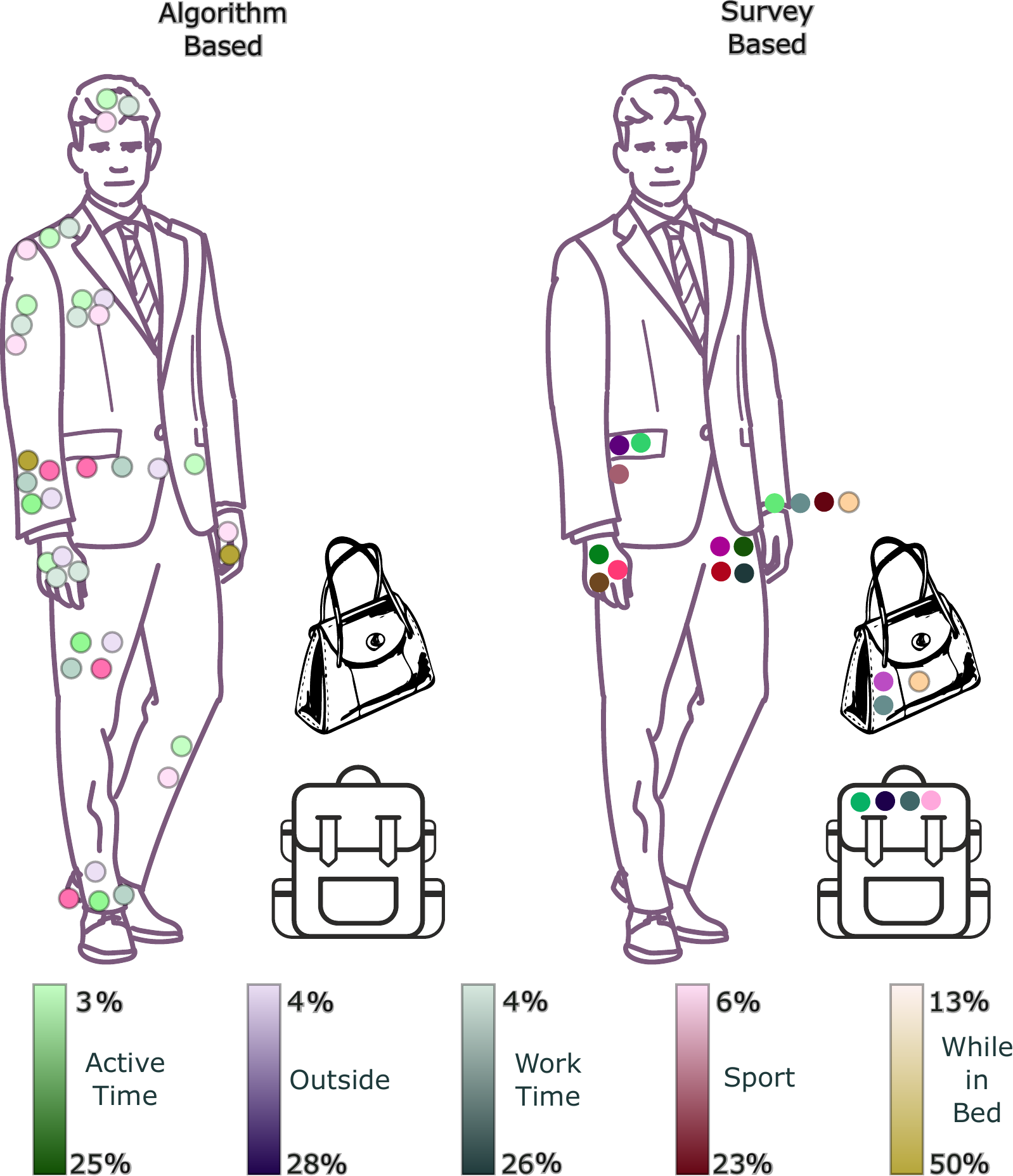}
    \caption{Comparison of optimal wearable placements identified by the~\citet{ray2025w2w} algorithm and placements reported by participants. Dots denote device locations; colour represents time of day and intensity indicates the percentage of responses. Percentages are computed relative to the total number of participant responses or algorithm location mentions.}
    \label{fig:position_comparison}
\end{figure}

To achieve this, we adopted a multi-step mapping process. First, we linked each reported wearing or carrying location from our survey to the corresponding wearable positions and interaction contexts identified in our synthesis of past guidelines (see~\autoref{sec:past_guidlines}), resulting in the mapping shown in~\autoref{tab:places_mapping}. Next, for comparison, we visualised both algorithmically recommended IMU sensor placements and user-reported device locations on human body silhouettes, generating comparative placement heatmaps (\autoref{fig:position_comparison}). As contemporary consumer wearables commonly incorporate IMU sensors, we aggregated placements across device types to support a unified comparison.

\begin{figure}
    \centering
    \includegraphics[trim= 0 0 0 0,clip,width=0.5\linewidth]{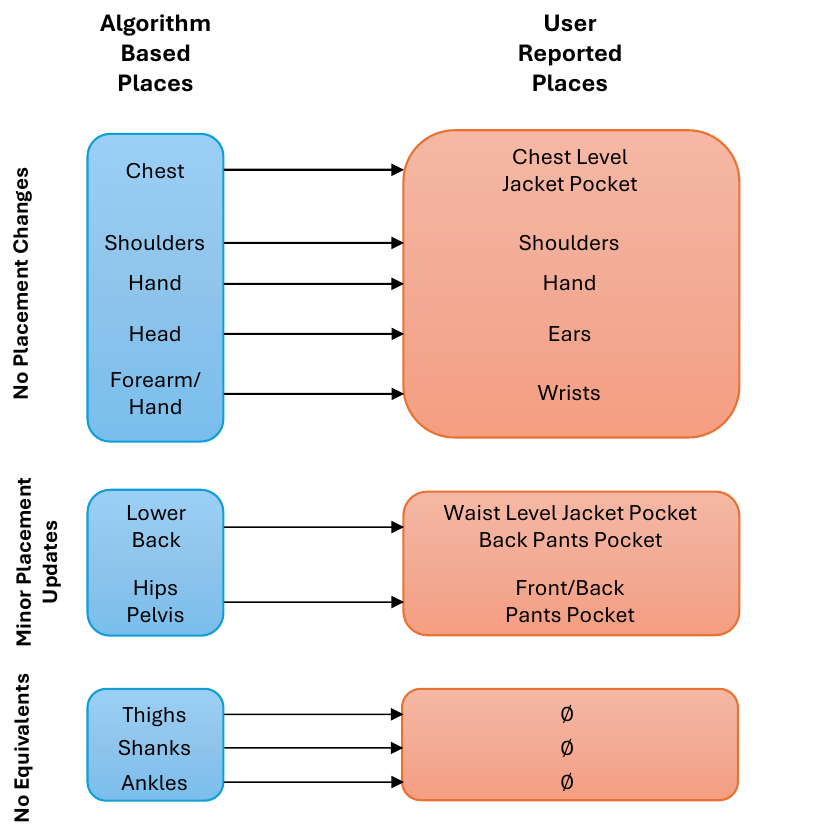}
    \caption{Sensor placements suggested algorithmically by \citet{ray2025w2w} compared to user-reported wearing locations of all devices combined.}
    \label{fig:old_new_mapping}
\end{figure}

To enhance interpretability and strengthen the connection between theory-driven recommendations and user-reported practices, we further aligned algorithmic placement categories with the bodily locations described by participants (\autoref{fig:old_new_mapping}). This additional mapping step enabled a more coherent comparison between technically optimal placement assumptions and the locations selected in everyday use. The resulting mappings and their detailed descriptions are presented below.

\begin{table}[htbp]
\centering
\caption{Mapping of the algorithm-based sensor placements as identified by~\citet{ray2025w2w}, based on task types and time contexts, as reported by participants in our sample.}
\label{tab:places_mapping}
\begin{tabular}{lll}
\toprule
\textbf{Time Context} & \textbf{Task} & \textbf{Body Areas Involved} \\
\midrule
Active time, & Locomotion & Thighs, Shanks, Lower Back \\
\cmidrule(lr){2-3}
Work hours,  & Object Manipulation & Forearms, Hands \\
\cmidrule(lr){2-3}
Sport       & Full Body Rotation & Pelvis, Chest, Ankles \\
\cmidrule(lr){2-3}
            & Communication Gestures & Forearms, Shoulders, Head \\
\cmidrule(lr){2-3}
            & Balance Intensive Tasks & Lower Back, Thighs, Ankles \\
\cmidrule(lr){2-3}
            & Asymmetric Tasks & Upper Arms, Hips \\
\midrule
Outside     & Locomotion & Thighs, Shanks, Lower Back \\
\cmidrule(lr){2-3}
            & Full Body Rotation & Pelvis, Chest, Ankles \\
\cmidrule(lr){2-3}
            & Object Manipulation & Forearms, Hands \\
\midrule
While in bed & Object Manipulation & Forearms, Hands \\
\bottomrule
\end{tabular}
\end{table}

\subsubsection{No Placement Changes}

\paragraph{Chest → Chest-Level Jacket Pocket}
Participants’ use of chest-level jacket pockets aligns spatially with the chest placement identified in algorithmic (\textbf{RQ3}). This correspondence is functional as well as anatomical—the chest pocket typically sits near the sternum or upper ribcage, regions that move rhythmically with respiration and body rotation. Devices in this position (such as smartphones or small sensors) can approximate chest motion data, though the stability and signal quality depend heavily on clothing fit. This placement also benefits from being discreet and easily reachable, supporting habitual use without discomfort or social awkwardness.

\paragraph{Shoulders → Shoulders (rarely, for Headphones)}
Although few participants reported direct device use at the shoulder level (\textbf{RQ3}), the rare occurrences were associated with headphone over-ear bands resting across the upper shoulders or neck region. From a biomechanical standpoint, the shoulders are relatively stable anchor points compared to distal extremities. Devices placed here could theoretically capture upper-body posture changes but would be limited in sensitivity. In real-world contexts, this location is seldom used for dedicated wearables due to comfort and visibility concerns, explaining the low frequency of reports.

\paragraph{Hand → Hand}
This mapping remains direct, as both the algorithmic and reported locations coincide (\textbf{RQ3}). Users frequently hold smartphones or other handheld devices, and this physical interaction corresponds exactly with algorithmic assumptions about hand-based motion signals. The hand is one of the most expressive and active parts of the body, making it an ideal source of fine-grained movement data. Its use, however, is intermittent—devices are picked up, manipulated, and set down—so while alignment between intended and actual placement exists, the temporal continuity of data may vary with behaviour.

\paragraph{Head → Ears}
No participant reported placing a device directly on the head itself (\textbf{RQ3}). However, the most anatomically and functionally adjacent location is the ears (\textbf{RQ3}). This mapping is straightforward, as earphones and earbuds are naturally positioned on or inside the ears—essentially an extension of the head. From a movement and signal perspective, the ears follow the head’s orientation and micro-movements closely, providing similar motion cues for activities like nodding or turning. Culturally and habitually, people are accustomed to wearing earphones throughout the day for calls, media, or passive listening, making them a convenient proxy for head-level sensing without the intrusiveness of a head-mounted device.

\paragraph{Forearm/Hand → Wrists}
Although the forearm and hand were indicated as relevant locations in previous algorithmic accuracy studies, users almost exclusively reported the wrist as their preferred and habitual placement of smartwatch (\textbf{RQ3}). This correspondence is anatomically logical, as the wrist is directly connected to both the forearm and the hand, moving as part of the same kinetic chain. The wrist provides excellent coupling for inertial data capturing arm movement, while being socially normalized through widespread smartwatch adoption. Unlike the forearm or back of the hand, the wrist is comfortable, secure, and compatible with continuous wear, which makes it a practical and socially accepted location for both tracking and interaction.

\subsubsection{Minor Placement Updates}

\paragraph{Lower Back → Waist-Level Jacket Pockets / Back Pants Pockets}
As presented in \autoref{sec:findings}, no user reported these exact anatomical areas directly (\textbf{RQ3}). However, nearby storage locations such as the waist-level of a jacket or the back pants pockets serve as functional substitutes (\textbf{RQ3}). The back pocket corresponds roughly to the posterior pelvic region, providing a stable but less dynamic representation of lumbar motion since it rests below the spine on the pelvis. The waist-level jacket pocket, while positioned at a similar vertical height, is on the anterior side of the body and can vary greatly depending on garment fit and closure type. This variability introduces potential signal attenuation or inconsistency. Both positions share proximity to the lower torso, making them plausible surrogates for lower-back-based sensing, though with certain trade-offs in motion fidelity.

\paragraph{Hips/Pelvis → Front/Back Pants Pockets}
The hips and pelvis are closely associated with the location of pants pockets, which makes this mapping anatomically coherent (\textbf{RQ3}). Both the front and back pockets move in synchrony with pelvic rotation during walking or standing transitions. This provides an indirect yet biomechanically valid proxy for pelvic motion capture. The high frequency of smartphone placement in these pockets reflects not just anatomical suitability but strong cultural and practical habits—pockets are secure, accessible, and habitual storage zones. Thus, even though the exact placement may differ slightly (anterior vs. posterior), both positions deliver comparable gross-movement signals related to gait and posture.

\subsubsection{No Real World Equivalents}

\paragraph{Thighs → —}
No participants reported storing or wearing devices directly on the thighs (\textbf{RQ3}). This is likely due to the absence of natural or convenient attachment points in this region, and the potential discomfort of movement interference. Although the thighs are critical for locomotion-related motion signals, users typically avoid strapping or clipping devices here. Clothing design also limits accessibility, as most pockets and accessories are positioned higher (pelvis level) or lower (shoes), leaving the thigh area underutilized despite its biomechanical relevance.

\paragraph{Shanks → —}
Similarly, no users reported placing devices on the shanks (\textbf{RQ3}). From an anatomical and movement perspective, the shank (lower leg) experiences high dynamic motion and frequent contact with clothing, making it unsuitable for stable device attachment. Without specialized gear (like sports bands or compression sleeves), this area lacks both comfort and practicality for everyday use. Consequently, despite algorithmic potential for precise gait analysis, real-world adoption in this region remains non-existent.

\paragraph{Ankles → —}
As noted in~\autoref{sec:findings}, only one participant mentioned using a device positioned close to the ankle—a Coros Pod attached to the shoe (\textbf{RQ3}). This represents an outlier rather than a trend. While the ankle and foot provide valuable biomechanical data for gait and running analysis, everyday device use in this region is hindered by accessibility and durability issues. Moreover, devices on shoes introduce additional noise from each step’s impact, complicating signal interpretation. Historically, few consumer products have targeted this placement, explaining the minimal adoption observed in the survey.

\section{Design Considerations for Usable Wearables} \label{sec:design_considerations}
Based on the findings, we distilled a few recommendations to remember about while creating a new wearable device or designing an experiment with users. 

%Here, we propose several design considerations for future wearable designs based on the findings from our survey.

\subsection{Design Consideration 1: Adjust Sensor Placement to the Time of Day and Activities Performed}
% As the\textit{Active Time}is describing literary the whole day when person is not in the bed, it might be challenging to pinpoint the single type of clothing or place as the best one for the sensor placement. For sure it would be left wrist with sensor in a smartwatch. When making a totally new device it might be difficult to make people adapt it on the left wrist as most of them wears there smartwatch. For the usability purposes, and taking clothing standards mentioned in previous section, second sensor could be a phone in a waist-level jacket pocket. Those places should be available in most cases, however, it is important to note that jacket is a weather/temperature dependent thing, and if it disappearing then back pants pocket could actually make better sens (just make sure that people are actually using pants in that time).

%\textit{Active Time}-> thighs, lower back, forearms, ankles | hand

% Work hours ->  thighs, lower back, forearms, ankles | hand

% Outside -> lower back, forearms, hands, pelvis | hand

% Sport ->  thighs, lower back, forearms, ankles |hand -> but that mostly will be probably during breaks

%\textit{While in bed}-> forearms, hands

Based on the mapping between algorithmically recommended body locations and the self-reported usage patterns of participants, it is possible to identify several practical configurations that balance signal quality, anatomical logic, and real-world usability (\textbf{RQ2}). The following paragraphs summarize potential optimal placements for different daily contexts, building upon both prior research recommendations and the contextual constraints discussed in the previous sections.

\paragraph{Active Time}
As the \textit{Active Time} category describes nearly the entire day when a person is not in bed, it is difficult to isolate a single ideal device location that would remain valid across all sub-activities. Nevertheless, \textbf{the wrist}—corresponding to the mapped forearm/hand area—emerges as the most consistent and reliable position, typically realized through a smartwatch. It provides continuous monitoring, comfort, and high adherence. For a secondary sensor, the \textbf{waist-level jacket pocket} or the \textbf{back pants pocket} (mapped from the lower back area) represents a practical and anatomically aligned position while being still in the range of human perception of their body (\autoref{sec:past_guidlines}) for a smartphone. Additionally, the \textbf{front pants pocket} can be considered an honorary mapping to the lower back region, as it follows a similar biomechanical logic: both front and back pockets move coherently with the pelvis and lumbar segment during locomotion, albeit from opposite sides of the body. Given its popularity and accessibility, the front pants pocket may even outperform back pocket usage in some contexts, despite minor differences in motion coupling.

Together, these positions—wrist and waist-level or pants pocket storage—cover a broad range of daily movements and remain socially acceptable and ergonomically feasible.

\paragraph{Work Hours}
During work hours, the optimal locations follow similar anatomical logic to \textit{Active Time} but are more strongly influenced by occupational norms and clothing constraints. \textbf{The wrist} remains the most robust option for continuous monitoring, as it is unobtrusive and compatible with most work environments. The second placement should balance accessibility and discretion. The \textbf{front and back pants pockets}, corresponding to the mapped pelvis and thigh regions, are the most practical for smartphones during work—especially in occupations requiring frequent device access or standing mobility. For settings with stricter policies or formal dress codes, a \textbf{waist or chest-level jacket pocket}, or desk/furniture surface may substitute as a temporary holding position.

Overall, the wrist and front/back pants pockets combination achieves a strong balance between functionality, accessibility, and signal continuity throughout typical work contexts.

\paragraph{Outside Activity}
In outdoor contexts, movement dynamics and environmental factors (temperature, clothing layers, body orientation) shape placement suitability. \textbf{The wrist} remains a consistent primary location, both for comfort and movement tracking. The secondary location should be stable yet accessible for quick interactions, such as navigation or communication. Based on the mapping, \textbf{the back pants pocket, waist-level jacket pocket, or front pants pocket} (all derived from lower back and pelvis mappings) serve this function effectively. These locations maintain sensor alignment with the trunk, allowing good capture of locomotion dynamics without obstructing movement. For active listening or calls, \textbf{earphones} (mapped from the head–ear relation) provide an additional functional data source, though not suitable as a stand-alone placement.

Hence, for outdoor scenarios, a combination of wrist and pocket-level placements yields optimal coverage of upper and lower body motion.

\paragraph{Sport and Exercise}
During sport or exercise, sensors must tolerate higher acceleration, sweat, and clothing movement. \textbf{The wrist}, corresponding to the forearm/hand area, continues to be the most stable and widely accepted location, especially due to smartwatch-based activity tracking. \textbf{The hand} may temporarily host a smartphone during short breaks or for direct control of media and applications, but it is not suitable for continuous sensing during movement.

For multimodal activity recording, combining the wrist sensor with a temporarily stored device at \textbf{waist pocket level} offers strong representational coverage without burdening the user or interfering with physical performance.

\paragraph{While in Bed}
In the resting context, the body assumes static or low-motion states, requiring placements that remain comfortable and safe during prolonged contact. \textbf{The wrist} continues to be ideal for continuous monitoring through smartwatch wear, particularly for sleep tracking applications, as it ensures signal consistency (e.g. heart rate, micro-movement) without disrupting comfort. However, the data also indicate a notably high frequency of smartphone use \textbf{in hand} while in bed. This behaviour is likely driven by active engagement in activities such as watching videos, reading news or books, messaging, or participating in calls before sleep. From a sensing perspective, such active hand use produces highly dynamic but short-duration motion data, distinct from passive placement scenarios. Once users transition to rest, the device is typically set aside on nearby furniture, such as a night stand or bed-side surface, corresponding to the habitual ``hand reach'' zone observed in the mapping. 

Therefore, the wrist remains the optimal continuous-wear position for physiological and movement monitoring, while the hand—and subsequently nearby furniture—represent realistic, behaviourally grounded placements for smartphone interaction and temporary rest states.

\begin{DesignConsideration}
    In summary, although temporal and activity contexts significantly influence user habits, wrist placement appears consistently across all the activities and time periods we analysed. Therefore, if the application allows, we recommend sticking to that placement. For other activities, we recommend careful consideration of stakeholders’ needs.    
\end{DesignConsideration}

\subsection{Design Consideration 2: No Uniform Device Placement across Users}
Our findings show that demographic factors, such as gender, age, education, and cultural influences, significantly shape how and where users carry wearable devices. For example, men predominantly use front pants pockets, while older users demonstrate more conservative and consistent placement patterns (\textbf{RQ1}).

% As public datasets often overrepresent young, male, STEM-affiliated participants~\cite{(NCSES)_2024}, overreliance on these datasets introduces biases when developing wearables or training Human Activity Recognition (HAR) models. However, collecting new data is also costly and time-consuming~\cite{liu2018benchmark, reiss2012creating}. %These limitations may lead to unrealistic assumptions. 

To date, researchers seeking to develop a new wearable device or train a Human Activity Recognition (HAR) or motion capture model typically face two options: (1) use an existing publicly available dataset, or (2) collect a new dataset. While collecting a new dataset allows for control over experimental conditions, it is costly and time-consuming~\cite{liu2018benchmark, reiss2012creating}. Recruitment is often restricted to student populations, particularly those in STEM disciplines, which introduces demographic biases such as skewed gender representation and age limitations~\cite{(NCSES)_2024}. Existing public datasets may appear more practical, but they frequently over-represent young, male, STEM-affiliated participants~\cite{(NCSES)_2024}, potentially introducing biases that limit the generalisability of wearable designs or HAR models.

These limitations can result in unrealistic assumptions about device placement.~\citet{leng2024emotion} designed an emotion recognition system using IMU sensors located in the front pants pocket, a location that is typically unsuitable for women. Similarly,~\citet{liu2022smart} proposed a wearable to monitor eating behaviour in older adults, yet lacked older participants in their study. Their chosen chest placement under the breast may also pose comfort issues for women. Another example includes a panic attack detection system with chest-worn sensors---effective in signal quality but potentially incompatible with women's undergarments~\cite{rubin2015towards}. Additionally, a system using headphone microphones for activity recognition may face adoption barriers, as many users now prefer smaller earbuds unsuitable for such hardware~\cite{yatani2012bodyscope}.

\begin{DesignConsideration}
To ensure real-world relevance, we recommend that future wearable designs consider diverse carrying preferences. 
Avoid defaulting to placements optimised for a limited subset of users, and instead design for flexibility and inclusivity across genders and age groups.
\end{DesignConsideration}
%Avoid defaulting to male-centric placements, and instead design for flexibility and inclusivity across genders and age groups.

\subsection{Design Consideration 3: Smartphone Availability During Sports and Throughout the Day}
Our findings show that smartphone placement varies significantly by gender and context. While men often use front pants or jacket pockets, women frequently store phones in purses or backpacks (\textbf{RQ1}). During physical activity or certain daily routines, users may not carry their phones at all, leading to gaps in sensing coverage (\textbf{RQ2}). This variability can hinder systems that assume consistent on-body smartphone availability.

\citet{leng2024emotion} have acknowledged such limitations due to reliance on specific body locations that are inaccessible or uncomfortable for some users. For example, the use of the front pants pocket---a common placement for smartphones---may hinder adoption, or introduce variability in algorithm performance when positioned elsewhere. The system proposed by~\citet{bian2024earable} addresses everyday activity recognition by utilizing a smartwatch and a custom earable, eliminating smartphone dependency. However, this approach introduces new challenges: the custom earable includes electrostatic sensors not found in standard earphones, which may hinder adoption. Users may prefer their own audio devices or find continuous wear uncomfortable, as noted in our survey. Another solution leverages a shoe-mounted device---commonly used in sports---for enhanced movement recognition \cite{hardegger2014enhancing}. While this setup prioritizes both algorithmic performance and user comfort during physical activity, its broader adoption may be limited. Our findings show that such devices (e.g. Coros Pod) are rarely used outside of sport-specific contexts and may be impractical for everyday or professional settings. Finally, authors of \cite{mirtchouk2016automated} acknowledge the limitations of their multi-sensor audio-motion system in everyday use, illustrating a broader issue: improvements in recognition accuracy often come at the cost of wearability and real-world practicality. This trade-off raises questions about the long-term adoption of such systems by the general public.

\begin{DesignConsideration}
We recommend avoiding the assumption that smartphones are always carried on the body. People often set their phones aside during physical activity, at home, at work, or in social situations, which can lead to data gaps or missed interactions. Systems that rely solely on smartphones for sensing may miss important moments. Complementary wearables can help ensure more consistent data collection across a wider range of everyday contexts.
\end{DesignConsideration}

\subsection{Design Consideration 4: Align with Users’ Dominant Wrist Preferences}
We found that most users wear smartwatches on their non-dominant wrist (typically the left), a consistent behaviour across age and gender groups driven by comfort, ease of use, and minimal interference with daily activities (\textbf{RQ1}). Systems that require wear on the dominant wrist can disrupt these habits, potentially lowering adoption, increasing discomfort, or degrading interaction.

\citet{scholl2015wearables} present a system designed for use in laboratory environments that requires a smartwatch to be worn on the dominant (right) wrist. This requirement may clash with users’ natural preferences and reduce compliance or comfort, especially during prolonged use. Similarly, the haptic feedback system introduced by \citet{villa2021assisting} as well as the indoor magnetic navigation system proposed by \citet{aiba2024magserea}, rely on a novel wrist-worn device, but may face adoption barriers unless it integrates with or supplements users’ existing smartwatch ecosystems. Our findings reinforce that users have established wrist preferences, and disregarding them can conflict with existing routines.

Energy-aware systems such as the one proposed in \cite{korpela2015energy} demonstrate a more adaptable approach by using wrist-mounted sensors compatible with smartwatch norms, though they do not explicitly consider wrist dominance. Other systems aimed at emotion tracking~\cite{wang2022emotracer, qin2020having, grzeszczyk2023decoding} leverage common smartwatch form factors and locations, reinforcing the importance of designing around existing user behaviour to ensure comfort, usability, and long-term engagement. Similarly, \citet{searle2021anticipatory} employed an existing smartwatch model for data collection, enabling the capture of real-world user data and supporting easy adoption by consumers who already own a smartwatch. \citet{li2022re} present a system that supports reconnection with nature and is deployed using a control unit and a sensor bracelet. While technical requirements may limit direct integration, the bracelet’s unobtrusive form factor and compact dimensions support comfortable everyday wear alongside a smartwatch.

\begin{DesignConsideration}
To promote adoption, wrist-based systems should respect dominant wrist-wearing habits, offer wrist-agnostic designs, or ensure compatibility with prevailing smartwatch ecosystems.    
\end{DesignConsideration}

\subsection{Design Consideration 5: Evolving User Preferences Driven by Fashion and Technology Trends}
User behaviour regarding wearable device placement is not static; it evolves alongside broader changes in fashion, lifestyle, and technological development (\textbf{RQ2}). Our findings indicate that while older users tend to exhibit consistent, conservative placement habits, younger users are more likely to experiment and adapt based on current trends (\textbf{RQ1}).

For instance, comparing our results to earlier studies reveals significant shifts. In 2007,~\citet{cui2007cross} reported that 14\% of men used a belt clip to carry their phone, whereas in our study, only 1.9\% reported doing so. Similarly, 11.5\% of women in 2007 carried their phones in hand or around the neck---today, 40\% report carrying phones in hand, and none around the neck. Device placement in skirts, once reported by 16.5\% of women, is now non-existent in our study. Some patterns remain relatively stable. For example, in both our study and the one by~\citet{zeleke2022mobile}, men frequently reported using front pants pockets while at work, outside, or in vehicles. However, even these patterns are subject to change with shifting device form factors, clothing styles, and cultural expectations. In this context, emerging sensing methods, such as conductive textiles~\citep{gowrishankar2025static}, offer a promising perspective by enabling sensing technologies to be integrated easily and unobtrusively into everyday clothing. Other approaches, including the use of cosmetics or makeup as sensing substrates~\citep{boldu2020maghair} and skin-compatible 3D-printed wearables~\citep{duente2024touch}, similarly support seamless integration into daily routines. However, because fashion trends are fluid, both the integration strategies and the algorithms operating on the resulting data must be readily adaptable to evolving styles.

\begin{DesignConsideration}
Therefore, designers should avoid relying solely on historical norms or current standards. Instead, they must consider how emerging trends and generational differences may influence user behaviour over time. Wearables should be designed with flexibility and adaptability in mind, ensuring long-term usability across diverse and evolving contexts.
\end{DesignConsideration}

\subsection{Design Consideration 6: Build Benchmarks that Reflect Real-World Device Placements}
Benchmarks and datasets used in wearable sensing research---especially in applications like HAR---often rely on fixed or researcher-assigned device placements (e.g. front pants pocket or wrist). While this simplifies data collection and model development, it fails to account for the wide variation in how people actually carry devices in everyday life. \citet{schlogl2015wearables} highlighted as early as $2015$ the need for more “in-the-wild” experiments, arguing that only such studies yield complete data. Our findings support this conclusion, showing that factors such as gender, age, fashion, and comfort significantly influence on-body device placement (\textbf{RQ1}, \textbf{RQ2}).

This mismatch between controlled benchmarks and real-world behaviour can lead to reduced performance or generalisability when algorithms are deployed outside lab settings. Fortunately, some existing benchmarks have already moved toward more valid practices. For instance,~\citet{sztylerOnBody2016} and~\citet{Fujinami2016OnBodySL} have proposed datasets that include multiple on-body positions, better reflecting users’ natural habits.

\begin{DesignConsideration}
To support fairer and more effective model evaluation, we recommend that future benchmarks prioritize capturing these real-world variations. This includes allowing participants to place devices where they typically would and annotating those locations. Benchmarks should also strive for demographic diversity to capture the full spectrum of carrying habits. Doing so will lead to more inclusive, realistic, and robust wearable sensing systems.    
\end{DesignConsideration}

\subsection{Limitations} % ways forward = future work

%DONE
% - in theory our study is larger scale = 300 people; but in reality if you want to analyse all of the demographic influences and connections then you need more people -> more people
    % Demographic skew: Possibly overrepresentation of tech-savvy, younger, or highly educated participants.
    % Cultural/geographic bias: Even with “international”, some regions may dominate responses.
% - we could not follow up on peoples answers toask them to specify what they mean 
% - Self-Reported Data Bias -> Memory bias and social desirability bias could affect responses (e.g. people reporting socially “normal” placements instead of actual behaviour)
% - Limited Contextual Accuracy -> Participants may interpret contexts (e.g. “work hours”, “active time”) differently.
% - Behaviour changes over weeks/months (e.g. seasonal clothing, device upgrades), but your study is likely cross-sectional. No validation of how stable people’s habits are over time

While our study was informed by prior guidelines and designed to address gaps in existing literature, several limitations should be acknowledged. Our dataset comprises 300 valid responses, which is sizeable for survey-based HCI research; however, this number remains insufficient for robust statistical testing across multiple intersecting demographic variables. When participant attributes such as country of residence, education level, and gender are combined, some subgroups are represented by only a small number of individuals, limiting the reliability of fine-grained comparisons. Nevertheless, our analyses capture general trends and offer indicative patterns for specific subgroups, which we believe provide a valuable starting point for future investigations.

% Our data also exhibits demographic and geographical biases. Although we sought to recruit participants with diverse educational and professional backgrounds, a substantial proportion of respondents were connected to university communities, reflecting our recruitment reach. Similarly, participant distribution was concentrated in countries geographically or socially closer to the authors, where recruitment networks were more accessible. To partially mitigate these biases, we enabled open-ended responses to capture a broader range of experiences and practices; however, these qualitative insights do not fully compensate for the lack of balanced representation.

Our data also exhibit demographic and geographic biases stemming from how participants were recruited. Although we aimed to include individuals with diverse educational and professional backgrounds, a substantial proportion of respondents were affiliated with university communities, reflecting the reach of our sampling approach. Participant distribution was likewise concentrated in countries geographically or socially closer to the authors, where existing networks were more accessible. As a result, the sample was designed to reflect the typical research population,, commonly described as WEIRD (Western, Educated, Industrialized, Rich, and Democratic)~\cite{henrich2010weirdest}, with only a few exceptions.

As with any self-reported, questionnaire-based study, our findings are subject to limitations related to recall accuracy and social desirability bias (present also in other fields \cite{benitez2004large, krohn2013explaining}). Participants may over- or under-estimate certain behaviours, particularly those that carry cultural, social, or normative connotations (e.g. perceptions of smartphone overuse). In addition, terms such as \textit{``during the day''} or \textit{``outside''}, while seemingly intuitive, may be interpreted differently across cultural contexts, potentially introducing variance in responses. To partially mitigate these limitations, we enabled open-ended responses to capture a broader range of experiences and practices; however, these qualitative insights do not fully compensate for the lack of balanced representation.

Finally, our study does not include a longitudinal component to observe how wearable habits evolve over time. Understanding temporal change is important, as participants may forget behaviours they do not engage in at the moment of reporting, and rapid shifts in technology ecosystems may influence wearing practices. Future longitudinal work would enable the community to distinguish between persistent behaviours and short-lived trends, offering deeper insight into how wearable habits develop and stabilise. While this was not the focus of our work, our findings do indicate behaviour shifts, visible through changes in design considerations.

Despite these limitations, our study provides a novel and empirically grounded contribution to understanding wearable placement practices %across diverse populations. 
among typical research participants, with respect to differences across fields, gender, and (limited) cultural backgrounds. We believe it serves as a valuable foundation for future research and supports ongoing efforts toward more inclusive and contextually aware wearable design.

% Mitigations

%-> do interviews, focused groups, more questionnaires
% - get more balanced sample in terms of age, occupation, education levels, and different countries 
    % - get more information how the culture influence how people are using wearable device -> types of clothing, social norms etc.

% Studies (e.g. REF) showcase that participants lack awareness of the potential health consequences of their mobile phone device usage. In addition, carrying habits can either reinforce or undermine such technology usage (e.g. REF). At the same time, accessible wearables benefits users in other manners, such as communication (REF), ..., and .... (REF). This raises the question to what extent the designs of such devices should be designed with the following trade-off in mind: ease of use over well-being. 
% At the same time, wearable designs should be accessible. Designing wearables both for accessibility and well-being---while also taking contextual factors and individual user preferences into consideration---presents a unique design challenge in HCI.
% This calls for more research on how wearables' design limitations influence both well-being and user experiences.
%-> maybe go more into depth about current ease of use / utility of wearable devices (or lack thereof), and the negative consequences of using these devices too much (e.g. making yourself dependent on it)
% -> solution (hence survey needed) is to understand contextual factors as we may want to choose utility in certain scenarios but maybe not in others, is there a way we can design for wellbeing by knowing this information?

\section{Conclusions}

%%%%already done
% - we've seen potential gaps in related work
% - we made a survey to get data to address information about those gaps
% - we compared our results with past guidelines
% - collected information about influence of a country, type of environment, gender, age, education, occupation etc. 
% - based on the comparison we made a set of recommendations
% - the results we have are challenging uniform placement of sensors

%%%%wrong section -> I think all of this should be talked about in the discussion, not conclusion
% - It is important to remember about social influence for a number of reasons
%     - for it to be adaptable by the end users
%     - making devices/algorithms/models that will use the type of data that can be realistically collected in the wild otherwise it does not make sense to make them
%     - when making futuristic devices -observe trends through a few last years among young people -> they eventually will be old enough to be the main customer base
%     - research wise -> when planning experiments remember that not everyone is wearing clothes everyday that have all of the pockets etc. that you want to use for sensor storing (e.g. pants with not big enough pockets, skirts) that people are wearing watches on non dominant hand so their perception of dominant hand generally is different etc.

% -there is still work to be done - more wider surveys, follow ups with participants to see actual reasonsetc.

Wearable technologies continue to expand in capability and prevalence, yet their effectiveness and long-term integration into everyday life depend on how people choose to wear, carry, and interact with them across diverse contexts. This work starts to address gaps in current HCI and wearable computing research concerning demographic, cultural, and behavioural influences on wearable placement habits by showcasing the habits of the research participant population vs the designs tested on those people. Despite the modest sample size, our findings already revealed clear differences in wearable placement habits, suggesting that a larger-scale study would likely uncover further diversity and nuance. By synthesising prior guidelines and conducting a multilingual survey across research channels in multiple countries, we contribute empirical evidence that challenges dominant design assumptions and highlights the contextual complexity underlying wearable use.

Our findings marked differences in wearable location practices across gender, age, lifestyle, and cultural background. For example, men predominantly store smartphones in front trouser pockets, whereas women report a wider range of storage alternatives, including handbags and backpacks. Smartwatches are primarily worn on the non-dominant wrist, indicating persistent adherence to traditional watch-wearing norms. By contrasting these findings with existing design recommendations, we propose context-informed considerations for sensor placement that better reflect everyday routines, social settings, activities, and temporal patterns of use.

This research advances the understanding of how user diversity shapes wearable interactions and highlights the need to account for heterogeneity in body-related practices. We encourage future work to expand the scope of inquiry through (i) more demographically and culturally diverse participant samples; (ii) longitudinal studies to examine changes in practices over time; and (iii) qualitative and mixed-method approaches to unpack the social, emotional, and contextual drivers underlying wearing choices. Addressing these directions will support the development of more adaptable and user-centred wearable systems capable of functioning across varied real-world settings.

Our work further informs design and research practice. We emphasise the importance of engaging with diverse user groups early in the design process, avoiding assumption-driven design based on personal experience or dominant user norms, and considering how everyday non-digital objects (e.g. traditional watches, accessories, clothing) can surface transferable insights for emerging wearable form factors. %By foregrounding demographic and contextual variability, this study provides a foundation for more inclusive, equitable, and socially attuned wearable design and lays the groundwork for future systems that better align with the realities of users’ lived experiences.
By foregrounding that even across the research participant population the demographic and contextual variability differ from assumed designs, this study provides a first step  for more inclusive, equitable, and socially attuned wearable design and lays the groundwork for future systems that better align with the realities of users’ lived experiences. 
\vspace{12pt}

\begin{Punchline}
\centering
    Ultimately, \textbf{wearable systems should adapt to people—not the other way around}.    
\end{Punchline}

\bibliographystyle{ACM-Reference-Format}
\bibliography{bibliograph}

\end{document}

%% file: tab_finding.tex
% \usepackage{booktabs}
\captionsetup{font=footnotesize}
\begin{table}[]
{\scriptsize
\caption{Results of the Chi-Square Test of Independence ($\chi^2$) with a p-value lower then 0.001, indicating strong significance (***).}
\label{tab:statistical_results}
\resizebox{\textwidth}{!}{
\begin{tabular}{@{}llll@{}}
\toprule
\textbf{Independent Var.}      & \textbf{Dependent Var.}      & \textbf{p-value}                & \textbf{Observations} \\ \midrule
Device                & Back Pants Pocket   & 9.60e-11 & Widely used for smartphone storage, less so for earphones, and rarely for smartwatches.       \\
Device                & Front Pants Pocket  & 1.18e-21 &  Generally used for smartphone storing.   \\
Device                & Backpack            & 3.72e-23 &  Earphones are the most commonly stored device in backpacks when not in use.        \\
Device                & In hand             & 7.10e-23 &  Smartphones are often held in hand, especially by people under 28.    \\
Device                & Waist Jacket Pocket & 7.29e-08 &  Typically used to store smartphones or earphones, rarely smartwatches.        \\
Device                & Left wrist          & 6.55e-287&  Smartwatches are usually worn by everyone, not stored.        \\
Device                & Purse               & 4.64e-07 &  Used to store various devices, primarily smartphones and earphones.        \\
Device                & In ears             & 1.03e-36 &  Only earphones are kept in ears while being used.      \\
Age                   & Back Pants Pocket   & 4.69e-07 & The back pocket is more common storage place for young people.         \\
Age                   & Front Pants Pocket  & 6.15e-04 &  Not the most popular among the youngest group (18-21) or those over 35.        \\
Age                   & Left wrist          & 2.26e-12 &  The older the person, the more likely they are to wear it on the left wrist.        \\
Age                   & Purse               & 1.66e-06 & Higher percentage of people after 35 use purse.        \\
Gender                & Back Pants Pocket   & 2.67e-08 &  The back pants pocket is more commonly used by women than men.       \\
Gender                & Front Pants Pocket  & 2.64e-57 & More frequently used by men than by women.         \\
Gender                & Backpack            & 2.31e-06 & Second most popular storage option among women.         \\
Gender                & Waist Jacket Pocket & 3.09e-06 & Third most popular among women overall.    \\
Gender                & Purse               & 4.53e-24 &  Generally used by women regardless of the device.     \\
Edu. lvl       & Back Pants Pocket   & 1.94e-05 &  Most popular among high school students.  \\
Edu. lvl       & Smartwatch *         & 7.88e-05 &  Highest usage reported by those with a master’s degree.        \\
Edu. lvl       & In ears             & 2.34e-06 &  Reported only by individuals with a master’s, bachelor’s, or practical edu..    \\
Env. Influence & Purse               & 8.48e-05 &  Most popular among people highly influenced by their environment (>75).     \\
Env. Influence & In ears             & 1.46e-08 &  Most reports come from those moderately influenced (50-75).        \\
Env. Influence & On Furniture        & 2.42e-06 &   Most users report being moderately influenced by their environment.    \\ \midrule
Device, Age                              & Front Pants Pocket  & 1.88e-16   &  Mostly popular for smartphones among younger people under 34.\\
Device, Gender                           & Back Pants Pocket   & 1.46e-19   &  Used by many women, though not their primary storage place.    \\
Device, Gender                           & Front Pants Pocket  & 5.36e-87   &  Mostly used by men, with no clear link to age or time.   \\
Device, Gender                           & Backpack            & 1.04e-27   &  Women store smartphones there at work and during sports.        \\
Device, Gender                           & In hand             & 4.92e-20   &  More women than men carry smartphones in hand. \\
Device, Gender                           & Waist Jacket Pocket & 4.70e-11   &  One of the storage options used by women. \\
%Device, Gender                           & Smartwatch          & 3.10e-277  &  Smartwatches are mainly worn on the left wrist.      \\
Device, Gender                           & Purse               & 1.80e-28   &  Women store smartphones there at work and during physical activities. \\
Device, Edu. lvl                  & Back Pants Pocket   & 3.80e-10   & Most popular for smartphones among people with practical or high school edu.        \\
Device, Edu. lvl                  & Front Pants Pocket  & 1.29e-17   &  Most popular among those with bachelor’s, master’s, or PhD degrees for smartphones and occasionally earphones.       \\
Device, Edu. lvl                  & Backpack            & 1.36e-23   &  Most common among “younger” degrees (below PhD), mainly for smartphones.      \\
Device, Edu. lvl                  & In hand             & 1.04e-18   &  Especially popular among “younger” degrees (below PhD), primarily for smartphones.    \\
Device, Edu. lvl                  & Waist Jacket Pocket & 2.95e-05   &  Used by all, but master's degree holders also often store earphones there along with smartphones.   \\
% Device, Edu. lvl                  & Smartwatch          & 4.49e-275  &  Most reports come from PhD and practical edu. holders, mostly for smartwatches, though some store smartphones.   \\
Device, Parents Influence                & Front Pants Pocket  & 2.12e-11   &  Those with the lowest parent influence report the highest use of this for both earphones and smartwatches.        \\
Device, Parents Influence                & Backpack            & 2.71e-09   & Highest parent influence → smartphones in backpack; lowest parent influence → earphones in backpack \\%The highest and lowest parent influence groups report the most usage—smartwatches for the highest, earphones (not smartphones) for the lowest.\\
Device, Parents Influence                & In hand             & 1.16e-11   &  The greater the parent influence, the more people report using their hand to hold a smartphone.   \\
% Device, Parents Influence                & Smartwatch          & 8.65e-221  &  People with lower parent influence report more smartwatch use on the wrist.  \\
Device, Friends Influence                & Back Pants Pocket   & 2.48e-05   &  Those extremely influenced or not influenced at all show the highest use of the back pants pocket for smartphone storage.        \\
Device, Friends Influence                & Front Pants Pocket  & 6.26e-15   &  People with the least influence report usage for all three device types.        \\
Device, Friends Influence                & Backpack            & 6.51e-16   &  Highly influenced individuals report the most usage; those with low influence report usage across all device types.        \\
Device, Friends Influence                & In hand             & 6.03e-14   &  The higher the parent influence, the more smartphones are held in hand and the fewer earphones are stored.     \\
% Device, Friends Influence                & Smartwatch          & 1.21e-282  &  People at both extremes of influence report the highest use of smartphone cases.  \\
Device, Friends Influence                & Purse               & 7.21e-09   &  Extreme influence levels are associated with the highest use of device cases overall.    \\
Device, Env. Influence            & Front Pants Pocket  & 1.15e-14   &  The higher the influence, the more people use it, with less variation across device types.     \\
Device, Env. Influence            & Backpack            & 4.37e-18   &  Higher influence is linked to greater use for both smartphones and earphones.    \\
Device, Env. Influence            & In hand             & 8.42e-17   &  With increasing influence, usage rises across multiple devices, including earphones.      \\
% Device, Env. Influence            & Smartwatch          & 1.73e-291  &  People with the least influence also use it for smartphone storage.   \\
Device, Env. Influence            & Purse               & 1.81e-08   & As influence increases, usage for both smartphones and earphones rises.   \\
Age, Gender                              & Back Pants Pocket   & 1.42e-17   &  The older the person, the less likely they are to use it for storage; women use this pocket more than men.       \\
Age, Gender                              & Front Pants Pocket  & 7.11e-54   &  Regardless of age, men prefer this spot for storing items.    \\
Age, Gender                              & Backpack            & 1.93e-05   &  Women use backpacks more than men; the older the group, the greater the gender difference.    \\
Age, Gender                              & Waist Jacket Pocket & 6.31e-08   &  In general, older men are the least likely to use backpacks.     \\
Age, Gender                              & Smartwatch          & 2.00e-11   &  Up to age 33 more women then men is wearing smartwatch. After age 33, more men report wearing smartwatch.  \\
Age, Gender                              & Purse               & 3.50e-21   & Women from  25-26, 31-39, 43+ age groups prefer to store smartphones and smartwatches in purses \\% Older women (43+) more commonly use a purse for storing smartphones and smartwatches, except for the 25-26 and 31-39 age groups, who also show purse use.\\
Gender, Edu. lvl                  & Back Pants Pocket   & 1.49e-09   &  More women use it overall, but among higher degree holders, more men use it.    \\
Gender, Edu. lvl                  & Front Pants Pocket  & 1.23e-51   &  Among those with practical edu., only men use it; men always outnumber women in usage.        \\
Gender, Edu. lvl                  & Backpack            & 3.43e-06   & More women than men use backpacks; among bachelor’s degree holders, most men use backpacks.  \\
Gender, Edu. lvl                  & Smartwatch          & 4.39e-06   & Only At PhD level men use smartwatch more than women \\% Most commonly used by master’s degree holders, with usage equal between genders; only among PhDs do more men use it than women.        \\
Gender, Edu. lvl                  & Purse               & 1.22e-23   &  Among high school and master’s degree, only women use it \\%; in other groups, both genders do, though only one male in some.   \\
Gender, Parent Influence                 & Front Pants Pocket  & 2.81e-37   &  Always more males use it; with increasing influence, female usage decreases.        \\
Gender, Parent Influence                 & Purse               & 8.39e-14   &  Always more women than men use it; in the 25-50 and 75-100 influence groups, no men report use.\\
Gender, Friends Influence                & Front Pants Pocket  & 7.17e-45   &  Always more men than women use it; usage appears unrelated to influence.  \\
Gender, Friends Influence                & Backpack            & 4.44e-08   &  Generally, women with higher influence are more likely to use backpacks; men show a similar pattern but to a lesser extent.
  \\
Gender, Friends Influence                & Purse               & 1.80e-20   &As influence increases, male usage declines.         \\
Gender, Env. Influence            & Back Pants Pocket   & 5.00e-05   &  Higher environmental influence leads to more backpack use among women; the same applies to men, but at a lower level.  \\
Gender, Env. Influence            & Front Pants Pocket  & 5.52e-44   & For both genders, greater influence leads to more usage; men still use it more than women overall.        \\
Gender, Env. Influence            & Backpack            & 3.13e-05   &  With rising influence, more women use backpacks; male usage decreases.   \\
Gender, Env. Influence            & Purse               & 3.58e-18   &  In general, those with higher influence use backpacks more. High schoolers and PhDs report less usage than other Edu. lvls.        \\
Edu. lvl, Friends Influence       & Waist Jacket Pocket & 5.04e-05   &  High schoolers with low influence use backpacks the most; master’s holders with moderate influence (50-75) use them the most.    \\
Edu. lvl, Friends Influence       & Smartwatch          & 2.30e-10   &  Master’s degree holders with extremely high or low influence report the highest usage.   \\
Parents Influence, Friends Influence     & Backpack            & 6.96e-07   & low parent influence \& low friend influence → less backpack use; high parent influence \& low friend influence → more backpack use\\%Among those with low parent influence (0-25), lower friend influence leads to less backpack use; among those with high parent influence (75-100), \\ 
%& & & lower friend influence leads to higher backpack use.\\
Parents Influence, Env. Influence & Purse               & 6.55e-09   &  
\{High Env. influence \& high parent influence\} or \{Low Env. influence \& low parent influence\} → less purse usage \\ %At the highest environmental influence level, higher parent influence reduces purse use; at the lowest environmental influence level, \\ 
%& & & lower parent influence reduces purse use.\\
Friends Influence, Env. Influence & Smartwatch          & 2.22e-04   &  Low Env. Influence (0-25): Higher Friend Influence → Increased smartwatch Use. \\
%& & & Moderate Env. Influence (50-75): Lower Friend Influence → Decreased Usage.\\ %Among those with the lowest environmental influence (0-25), higher friend influence increases left wrist use; for those with moderate environmental influence (50-75), \\ lower friend influence leads to less usage.
Friends Influence, Env. Influence & Purse               & 1.07e-13   &  \{High Env. Influence \& low Friend Influence\} or \{low Env. Influence \& high Friend Influence\}  → Increased Purse Use. \\ 
% & & & Low Env. Influence (0-25), High Friend Influence (75-100): Maximal Purse Use. %People with high environmental influence (75-100) and low friend influence (\textless{}50) tend to use purses more than others; those with low environmental influence (0-25)\\  use purses most when highly influenced by friends (75-100). 
\midrule
Device, Gender, Friends Influence        & Front Pants Pocket  & 4.18e-64   & Males with low friend influence and women with very low friend influence  → more smartwatch storage in front pockets \\ 
%& & &, who store their smartphones there. While still fewer than men, this group includes slightly more women than other influence levels.\\
Device, Gender, Env. Influence    & Front Pants Pocket  & 2.64e-64   & Males with higher Env. influence → more smartphones storage in front pockets; Males with moderate Env. influence → \\
& & & more earphone storage in front pockets; Women with high Env. influence → more smartphone/earphone storage in front pockets\\% with higher influence, but  muchs lower than men’s. \\
%For males storing smartphones, the higher the influence, the more frequently they use the front pants pocket. Men with moderate environmental influence \\ 
%& & & store earphones in the front pants pocket more than others. For women, higher influence leads to increased use of the front pants pocket for smartphones \\ 
%& & & or earphones, though their usage remains much lower than that of men. \\
\bottomrule \\
\multicolumn{4}{l}{*We found the left wrist demonstrated near-exclusive use for smartwatches among our participant cohort, thus in the following combinations, we replaced left wrist with smartwatch.} 
\end{tabular}
}
}
\end{table}